# Dynamic Flow Equilibrium of Transportation and Power Distribution Networks Considering Flexible Traveling Choices and Voltage Quality Improvement

Jiaqi Li, Xiaoyuan Xu, *Member, IEEE*, Zheng Yan, *Member, IEEE*, Han Wang, and Yue Chen

***Abstract*—The increasing global spread of electric vehicles (EVs) has introduced significant interdependence between transportation and power networks. Most of the previous studies on coupled networks focus on the formation of equilibrium states between transportation and distribution networks but neglect diminishing the adverse influences of the interaction, especially the voltage quality problems caused by large and fluctuating charging loads. This paper constructs a dynamic interaction model of coupled networks considering both economic and safe operation. First, the dynamic traffic assignment problem is established considering both departure time and route choices, and a dynamic queuing model is designed to describe the EV queuing behaviors at fast charging stations (FCSs). Then, the charging load fluctuation toll is designed as an ancillary toll at FCSs and the bus voltage deviation penalty is considered in optimal power flow to guide the interaction of two networks and tackle the voltage deviation issue. Moreover, the equilibrium state is designed as a fixed-point problem and the solution's existence is proved under mild assumptions. Third, the linearization and convex relaxation techniques are used to improve computational efficiency and a Monte Carlo simulation technique is developed to evaluate the influence of uncertain travel demands on coupled networks. Numerical simulations of coupled networks interaction analyses in deterministic and uncertain conditions are presented.**

## I. Introduction

ENERGY and transportation have always been two core issues in the development of human society. Recently, the fossil resource shortage and environmental pollution bring severe challenges to the traditional transportation industry [1-2]. With policy support and technology advances, the market share of electric vehicles (EVs) has increased rapidly and the stock is predicted to achieve 2450 million by 2030 [3]. As an indispensable facility to meet long-distance EV travel, the fast charging station (FCS) gradually plays a vital role in the intelligent transportation network (TN). Nowadays, the rated power of superchargers is up to 100kW [4], and the considerable charging demand at FCSs has affected the operation of the distribution network (DN) [5-7]. The increasing spread of EVs is expected to introduce significant interdependence between two systems. In TN, the charging prices affect the travel choices of EVs, which determines the distributions of charging load. In DN, the charging demands affect the optimal operation state, which further impacts the charging prices. It is crucial to explore interaction mechanisms and develop coupled models for joint flow analysis, coordinated operation and integrated planning of the two systems [8].

The modeling of coupled transportation and power networks mainly concerns three aspects: interaction, TN and DN models. To our knowledge, reference [9] is the first work shedding light on the equilibrium state between TN and DN, where the retail electricity price is designed as the locational marginal price (LMP) and DC optimal power flow (OPF) is used for power market clearing. In the follow-up studies, LMP is usually treated as the interaction signal for the coupled networks [9-10]. As to DN models, considering that FCSs are connected to low-voltage power systems with a high R/X ratio, ACOPF has been used to model the economic operation of DN [10-14]. Besides, multi-period ACOPF has also been discussed in [16-21] to describe variable power system states.

Establishing TN models is the key task in investigating the interaction between TN and DN. The traffic assignment problem (TAP) is prevalent for obtaining the traffic flow distributions. Based on the Wardrop's first principle, the traffic flow distributions spontaneously reach a user equilibrium (UE) state owing to the selfishness of individual travelers, which means that the travel times on all used routes are equal and minimal [15]. The UE conditions are formulated as a mathematical program with complementary constraints (MPCC) [9] or converted to a convex optimization problem [10]. In addition, the social optimum (SO) is the other principle to achieve the equilibrium state of TN, where a central operator regulates all travels to minimize the total travel cost. The above works are based on the static traffic assignment (STA) models,

This work was supported by National Natural Science Foundation of China (U2166201, 52107116).
Jiaqi Li, Xiaoyuan Xu (corresponding author), Zheng Yan and Han Wang, are all with the Key Laboratory of Key Laboratory of Control of Power Transmission & Conversion and Shanghai Non-carbon Energy Conversion & Utilization Institute, Shanghai Jiao Tong University, Shanghai 200240, China (e-mail: ljq0324@sjtu.edu.cn; xuxiaoyuan@sjtu.edu.cn yanz@sjtu.edu.cn; wanghan9894@sjtu.edu.cn).
Yue Chen is with the Department of Mechanical and Automation Engineering, the Chinese University of Hong Kong, Hong Kong 999077, SAR China (e-mail: yuechen@mae.cuhk.edu.hk).

which assume that trips are completed instantly. However, the departure and travel times vary for different vehicles. Compared with STA, the dynamic traffic assignment (DTA) is preferable to describe the spatial and temporal variations of traffic flow. Moreover, as the DN operating states are time-varying, it is more appropriate to use DTA in analyzing coupled networks. In [16-17], the semi-dynamic traffic assignment (SDTA) model is established to describe the traffic flow propagation between adjacent periods. However, STDA only allows the residual traffic to transfer to the next period; thus, it is coarse for short-term operation and control. In [18-19], the DTA model based on the link performance function is exploited for the time-varying travel demand. To capture the forward and reverse queuing waves, the bi-directional wave dynamic model is used in [20] to characterize the traffic flow variation in coupled networks. In [21], a dynamic traffic model with point queues is developed, and the dynamic network equilibrium encapsulates the driver's choices of route, departure time, charging locations, and electricity price. In summary, DTA provides detailed and accurate traffic flow, and it deserves wide attention in studying the dynamic interaction between TN and DN. However, the existing works on dynamic coupled networks models mainly focus on the mechanism and formation of the equilibrium state between TN and DN, and do not design suitable control strategies to diminish the adverse influence of interaction. Due to the high-rated power of superchargers, the large and variable charging loads at FCS significantly affect the voltage quality of DN. Hence, it is important to guide the interaction behaviors of TN and DN considering the economic and safe operation of the two systems. In addition, the computational tractability is the main issue in establishing the dynamic coupled networks model and the solution algorithms are worth further exploration.

Furthermore, the uncertainty in such coupled networks is ubiquitous. The traffic demands in [22] are obtained via Monte Carlo simulation (MCS), and a stochastic optimization method is proposed for FCS planning. In [23], by solving STA in extreme scenarios, the traffic demand variation is mapped into the power demand uncertainty, and a two-stage robust optimization model is proposed for coupled operation. Reference [24] uses the travel demand scenarios to obtain the distributions of traffic densities for the wireless charging system planning. However, the influence of uncertainty on the equilibrium state of coupled networks has not been highlighted in the existing works.

This paper proposes an efficient and practical model for the economic and safe operation of coupled networks with flexible traveling choices. The penalties of bus voltage deviation and charging load fluctuation are considered to eliminate the adverse influence of large and variable load demands of FCS. Besides, the uncertainty analysis techniques are used to investigate the stochastic equilibrium states of coupled networks. The main contributions are summarized as follows:

1) The multi-period TN and DN models are established to describe time-varying traffic flows and distribution system operating states, respectively. The dynamic user equilibrium (DUE) conditions are used for DTA with both departure time and route choices. Moreover, considering the variation of queue length at FCS, a dynamic queuing model based on the dynamic balance between EV charging demands and idle charging piles is designed to capture detailed queuing time.

2) The penalties of bus voltage deviation and charging load fluctuation are considered in DN and TN to diminish the adverse influence of coupled networks interaction, especially the voltage quality problems caused by remarkable charging load demands. The former one is added to the objective function of OPF, which directly affects the DN operation and indirectly guides the traffic flow through LMP. The latter one, which is designed as an ancillary toll at FCS, affects the charging choices of EVs and reduces the impact of load fluctuation on DN. Based on the guidance of interaction, the equilibrium state of the economic and safe operation of coupled networks is designed as a fixed-point problem, of which the solution's existence is proved under mild assumptions.

3) The piecewise linearization and second-order cone relaxation techniques are utilized to improve computational efficiency in analyzing the coupled networks, and a decomposition algorithm is designed to solve the TN and DN sub-problems iteratively. Finally, the influences of uncertain travel demands on both TN and DN are analyzed via the MCS technique, which reveals the stochastic states of TN and DN in uncertain environments. Numerical experiments show that the proposed method obtains the equilibrium states of TN and DN efficiently and accurately.

The rest of the paper is organized as follows. Section II presents the DTA model. The equilibrium state of the coupled networks boils down to a fix-point problem in Section III, followed by the solution algorithm in Section IV. Section V gives the case studies and Section VI concludes our work.

## II. Transportation Network

This section presents the mathematical model and solution algorithm for the DTA problem.

### A. Network Loading Model

The transportation system is represented as the graph $G_T = [T_N, T_A]$, where $T_N$ and $T_A$ denote the node and link sets, respectively. Let $r, s \in T_N$ represent the origin and destination nodes, respectively, and $a \in T_A$ represents the road of a transportation network. Then, the path connecting the origin-destination (O-D) pair $(r, s)$ is represented by $p \in P^{rs}$.

#### 1) Flow Conservation Constraints

The flow conservation constraints depict the spatial variation of traffic flow, as illustrated in Fig. 1. The relationship between inflow $u_a(t)$, traffic flow $x_a(t)$ and exit flow $v_a(t)$ of link $a$ at time $t$ is expressed as:

$$\frac{\mathrm{d}x_{ap}^{rs}(t)}{\mathrm{d}t} = u_{ap}^{rs}(t) - v_{ap}^{rs}(t) \qquad \forall a, p, (r,s) \tag{1}$$

$$\sum_{p,rs} u_{ap}^{rs}(t) = u_a(t), \sum_{p,rs} x_{ap}^{rs}(t) = x_a(t), \sum_{p,rs} v_{ap}^{rs}(t) = v_a(t) \qquad \forall a \tag{2}$$

where $\mu_{ap}^{rs}(t)$, $x_{ap}^{rs}(t)$, $v_{ap}^{rs}(t)$ are the inflow, traffic flow and exit flow of link $a$ through path $p$ between O-D pair $(r, s)$ at time $t$, respectively. In general, the inflow $u_a(t)$ is regarded as a control variable, and $x_a(t)$, $v_a(t)$ are treated as state variables.

The flow conservation at node $j$ ($j \neq r, s$) indicates that the flow exiting from the links traveling towards node $j$ is equal to that entering the links departing from node $j$, which is stated as:

$$\sum_{a\in A(j)} u_{ap}^{rs}(t) = \sum_{a\in B(j)} v_{ap}^{rs}(t) \qquad \forall p,(r,s), j \neq r,s \tag{3}$$

where $A(j)$ and $B(j)$ are the sets of links whose starting and ending nodes are $j$, respectively.

For the origin node $r$ and destination node $s$, the flow conservation constraints are:

$$\sum_{p}\sum_{a\in A(r)} u_{ap}^{rs}(t) = \sum_{p} f_p^{rs}(t) = f^{rs}(t) \qquad \forall p,(r,s) \tag{4}$$

$$\sum_{p}\sum_{a\in B(s)} v_{ap}^{rs}(t) = \sum_{p} e_p^{rs}(t) = e^{rs}(t) \qquad \forall p,(r,s) \tag{5}$$

where $f^{rs}(t)$ and $e^{rs}(t)$ are the instantaneous departing and arriving rates between O-D pair $(r, s)$ at time $t$, respectively. The total travel demand $F^{rs}$ during the period $[0, T]$ is:

$$F^{rs} = \int_0^T f^{rs}(t)\mathrm{d}t \qquad \forall (r,s) \tag{6}$$

and $E^{rs}(t)$ represents the cumulative traffic flow arriving at destination $s$ by time $t$, as stated below:

$$E^{rs}(t) = \int_0^t e^{rs}(\omega)\mathrm{d}\omega \qquad \forall (r,s) \tag{7}$$

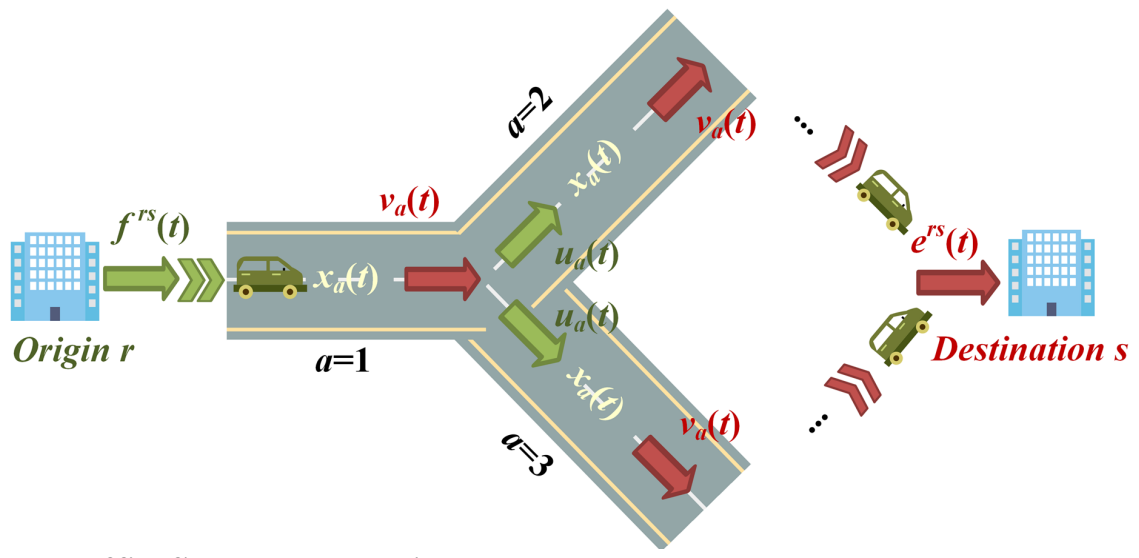

Fig. 1. Traffic flow conservation.

### 2) Flow Propagation Constraints

To reflect the temporal flow propagation, the dynamic traffic model is used in this paper and the following path-based flow propagation constraint is established:

$$x_{ap}^{rs}(t) = \sum_{b\in p'} \left\{ x_{bp}^{rs}[t+\tau_a(t)] - x_{bp}^{rs}(t) \right\} + \left\{ E_p^{rs}[t+\tau_a(t)] - E_p^{rs}(t) \right\} \qquad \forall a,p,(r,s) \tag{8}$$

where $p'$ is the sub-path of path $p$ from the end of link $a$ to the destination $s$, and $\tau_a(t)$ is the travel time of link $a$ at time $t$. Equation (8) shows that the vehicles on link $a$ at time $t$ will result in the extra vehicles on downstream links of the sub-path and the increased vehicles arriving at the destination at time $t + \tau_a(t)$.

### 3) Boundary Constraints

All the variables are non-negative, as stated below:

$$\begin{array}{lll} u_{ap}^{rs}(t) \geq 0 & v_{ap}^{rs}(t) \geq 0 & x_{ap}^{rs}(t) \geq 0 \\ f_p^{rs}(t) \geq 0 & e_p^{rs}(t) \geq 0 & \forall a,p,(r,s) \end{array} \tag{9}$$

and the initial values of $x_{ap}^{rs}$ and $E^{rs}$ are equal to zero:

$$x_{ap}^{rs}(0) = 0 \qquad E^{rs}(0) = 0 \qquad \forall a,p,(r,s) \tag{10}$$

## B. Travel Choice Principle

In most of the previous researches on coupled networks, only the route choice was considered in the DUE condition. To fully describe the travel choices, both the departure time choice and route choice are considered in this paper, as stated below:

$$\begin{cases} c_p^{rs*}(t) - c_{\min}^{rs*} \geq 0 \\ f_p^{rs*}(t)\left[c_p^{rs*}(t) - c_{\min}^{rs*}\right] = 0 & \forall t,p,(r,s) \\ f_p^{rs}(t) \geq 0 \end{cases} \tag{11}$$

where the superscript asterisk denotes a feasible solution, $c_p^{rs*}(t)$ is the travel cost for path $p$ at time $t$ between O-D pair $(r, s)$, and $c_{\min}^{rs*}$ is the minimal travel cost between O-D pair $(r, s)$. Based on (11), for each O-D pair, the travel costs of travelers, regardless of the departure time, are equal and minimal [25]. The equivalent variational inequality (VI) formula of (11) is given as (12.1):

$$\sum_{a}\sum_{t} c_a^*(t)\left[u_a(t) - u_a^*(t)\right] \geq 0 \tag{12.1}$$

$$c_p^{rs}(k) = \sum_{a}\sum_{t} c_a(t)\delta_{apk}^{rs*}(t) \qquad \forall p,k,(r,s) \tag{12.2}$$

$$u_a(t) = \sum_{r,s}\sum_{p}\sum_{k} f_p^{rs}(k)\delta_{apk}^{rs*}(t) \tag{12.3}$$

where $c_a(t)$ is the travel cost of link $a$ at time $t$, $c_p^{rs}(k)$ is the travel cost of path $p$ between O-D pair $(r, s)$ at time $k$. When the traffic flow between $(r, s)$ departing at time $k$ through path $p$ traverses on link $a$ at time $t$, $\delta_{apk}^{rs*}(t)$ is 1; otherwise, $\delta_{apk}^{rs*}(t)$ is 0.

## C. Solution Algorithm of DTA

The DTA problem considering departure time and route choices is equivalent to finding a vector $u_a^*(t)$ satisfying (12) and traffic flow constraints (1)-(10). The travel cost $c_a(t)$ in (12) is the travel time $\tau_a(t)$ multiplied by the monetary value $\omega$. Generally, the travel time $\tau_a(t)$ is expressed as:

$$\tau_a(t) = f\left[u_a(t), x_a(t), v_a(t)\right] \qquad \forall a \tag{13}$$

As shown in (8), the traffic flow only relates to downstream vehicles in DTA. Therefore, both $x_a(t)$ and $v_a(t)$ are substituted by the inflow $u_a$, and the travel time $\tau_a$ is described as a function of $u_a$ as follow:

$$\tau_a(t) = f\left[u_a(t), u_a(t\text{-}1), \ldots, u_a(1)\right] \qquad \forall a \tag{14}$$

The Jacobi matrix of (14) is asymmetric due to the link flow interaction; thus, the link-based VI (12.1) cannot be converted into an equivalent optimization problem. Here, we utilize the nested diagonalization method to solve the DTA problem. In addition, the time is divided into $T_{\mathrm{TN}}$ intervals and the discrete-time DTA model is used to analyze the interaction between TN and DN.

According to the flow propagation constraint (8), there are two sources of link flow interactions: 1) the travel time $\tau_a(t)$ that varies with traffic flow and needs to be estimated as $\overline{\tau}_a(t)$; 2) the interference among link inflows according to flow propagation with $\overline{\tau}_a(t)$. Here, the nested diagonalization method is used to eliminate link flow interactions. As shown in **Algorithm 1**, the travel time is first relaxed by the outer diagonalization method to yield a sub-problem. Then, in the diagonalization solution procedure, the interference among inflows is further relaxed to yield a tractable optimization problem.

*Remark 1:* In step **3**, the estimated travel time is rounded to the closest integers because of time discretization. Note that the estimated time is only used in flow propagation constraint (8) and it does not affect the objective function (16).

*Remark 2:* In step **6**, when the link travel time is temporarily fixed as $\overline{\tau}_a(t)$ ,the relationships among inflow, exit flow, and the number of vehicles are determined through the flow propagation constraint (8), and all the discrete constraints (1)-(10) are linear.

**Algorithm 1:** Nested diagonalization method

1 **Input:** $n$=0, $\overline{\tau}_a^n(t)=$ initial value $\forall a,t$

2 **repeat** // Outer loop

3 Update link travel time:

$$\overline{\tau}_a^n(t)=i \quad \text{if } i-0.5\le \tau_a^n(t)<i+0.5 \quad \forall a,t \quad (15)$$

4 **repeat** // Inner loop

5 $m$=1, $u_a^m(t)$ = initial feasible solution

6 Solve optimization problem:

$$\min f_{TN}=\omega\sum_t\sum_a\int_0^{u_a^{m+1}(t)}\tau_a(\theta,u_a^m(t-1),...,u_a^m(1))\mathrm{d}\theta \quad (16)$$

s.t. Discrete (1)-(10), $\tau_a(t)=\overline{\tau}_a^n(t)$ $\forall a,t$ in (8)

7 Inner loop index: $m$=$m$+1

8 **until** $\dfrac{\left\|\boldsymbol{u}_a^m(t)-\boldsymbol{u}_a^{m-1}(t)\right\|_2}{\left\|\boldsymbol{u}_a^{m-1}(t)\right\|_2}<\varepsilon_1$

9 Outer loop index: $n$=$n$+1

10 **until** $\dfrac{\left\|\boldsymbol{\tau}_a^n(t)-\boldsymbol{\tau}_a^{n-1}(t)\right\|_2}{\left\|\boldsymbol{\tau}_a^{n-1}(t)\right\|_2}<\varepsilon_2$

11 **Output:** $u_a(t)$, $\tau_a(t)$

## III. Coupled Traffic and Power Networks

In this section, we first expand TN with FCSs and develop a dynamic queuing model to describe the variable queuing time. The load fluctuation toll is considered at each FCS to reduce the impact of charging load fluctuation on DN. Then, we design the bus voltage deviation index in the multi-period OPF problem based on the branch flow model (BFM) for DN. Finally, we formulate the equilibrium of coupled TN and DN as a fixed-point problem and discuss the existence and uniqueness of its solution.

### *A. Transportation Network with FCSs*

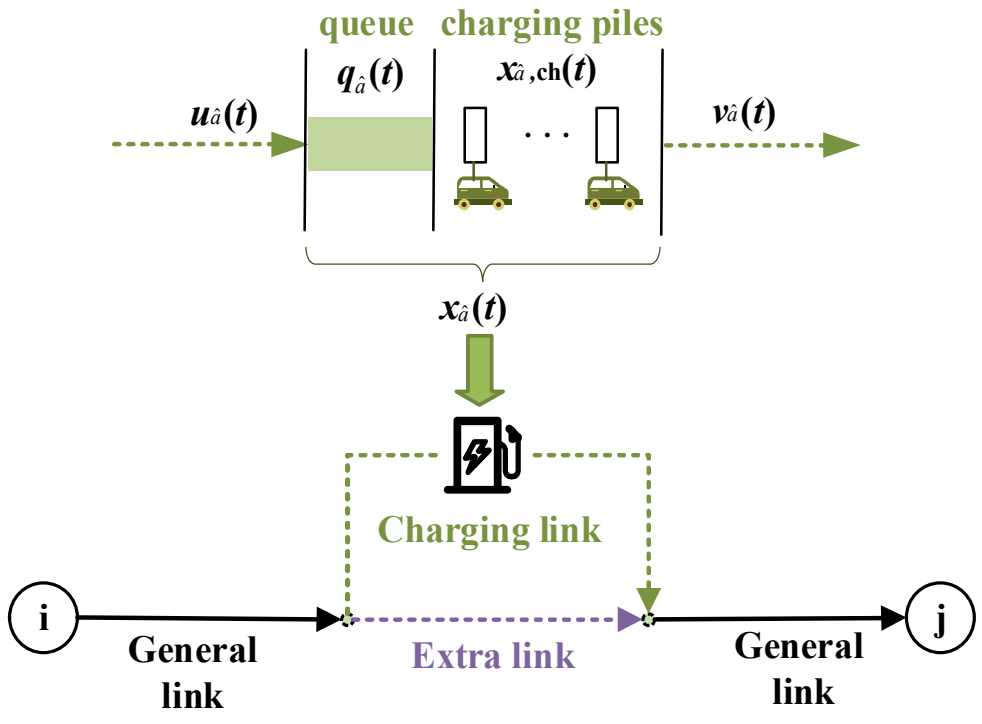

Fig. 2. Extended graph and dynamic queuing model of FCS.

In Fig. 2, the link of TN is augmented with two virtual links to distinguish the travel and charging choices of different vehicles, and the dynamic queuing model of FCS is shown. The expanded graph is stated as $\hat{G}_{\mathrm{T}}$=[$\hat{T}_{\mathrm{N}}$, $\hat{T}_{\mathrm{A}}$] .

*1) Charging link*

An EV that chooses to get charged will go through the charging link, of which the time cost is:

$$\tau_{\hat a}(t)=\tau_{\mathrm{ch}}+\tau_{\hat a,\mathrm{que}}(t)=\tau_{\mathrm{ch}}+\frac{q_{\hat a}(t)}{cap_{\hat a}}\cdot\tau_{\mathrm{ch}} \qquad \forall \hat a\in\hat{T}_{\mathrm{A,ch}} \quad (17)$$

where $\hat a$ is the link of expanded TN, $\hat{T}_{\mathrm{A,ch}}$ is the set of charging links, $\tau_{\mathrm{ch}}$ is the constant charging time, $\tau_{\hat a,\mathrm{que}}(t)$ is the dynamic queuing time relevant to the queue flow $q_{\hat a}(t)$ and FCS capacity $cap_{\hat a}$ . The EV queuing is determined by the number of idle charging piles:

$$q_{\hat a}(t)=x_{\hat a}(t)-x_{\hat a,\mathrm{ch}}(t) \qquad \forall \hat a\in\hat{T}_{\mathrm{A,ch}} \quad (18)$$

$$x_{\hat a,\mathrm{ch}}(t)=\min\{cap_{\hat a},x_{\hat a}(t)\} \qquad \forall \hat a\in\hat{T}_{\mathrm{A,ch}} \quad (19)$$

where $x_{\hat a,\mathrm{ch}}(t)$ is the charging flow of link $\hat a$ at time $t$. Then, the dynamic queue flow is formulated as

$$q_{\hat a}(t)=\max\{0,x_{\hat a}(t)-cap_{\hat a}\} \qquad \forall \hat a\in\hat{T}_{\mathrm{A,ch}} \quad (20)$$

Therefore, the travel cost is

$$c_{\hat a}(t)=\omega\tau_{\hat a}(t)+\lambda_{\hat a}(t)P_{\mathrm{ev}}\Delta t+T_{\hat a}^{\mathrm{L}} \qquad \forall \hat a\in\hat{T}_{\mathrm{A,ch}} \quad (21)$$

where $\lambda_{\hat a}(t)$ is the charging price of FCS at time interval $t$, $P_{\mathrm{ev}}$ is the EV charging power, and $\triangle t$ is the length of time interval. The large charging load fluctuation of FCS threatens the operation safety of DN; thus, FCS is charged with a load fluctuation toll $T_{\hat a}^{\mathrm{L}}$ :

$$T_{\hat a}^{\mathrm{L}}=\alpha\frac{\sigma_{\hat a}^{\mathrm{L}}}{\mu_{\hat a}^{\mathrm{L}}} \qquad \forall \hat a\in\hat{T}_{\mathrm{A,ch}} \quad (22)$$

where $\sigma_{\hat a}^{\mathrm{L}}$ and $\mu_{\hat a}^{\mathrm{L}}$ are the standard deviation and mean of the charging load in the charging link $\hat a$ ,respectively, and $\alpha$ is the monetary value.

*2) Extra link*

The Gasoline vehicles (GVs) and EVs that choose not to get charged traverse on the extra link, where the travel cost is zero.

$$c_{\hat a}(t)=\omega\tau_{\hat a}(t)=0 \qquad \forall \hat a\in\hat{T}_{\mathrm{A,ex}} \quad (23)$$

where $\hat{T}_{\mathrm{A,ex}}$ is the set of extra links.

*3) General link*

The travel time of general links is expressed as the latency function according to the modified Bureau of Public Roads model:

$$\tau_{\hat a}(t)=\tau_{\hat a}^0\left[1+\left(\frac{x_{\hat a}(t)}{cap_{\hat a}}\right)^{\beta}\right] \qquad \forall \hat a\in\hat{T}_{\mathrm{A,gen}} \quad (24)$$

$$\beta=b+c\cdot\left(\frac{x_{\hat a}(t)}{cap_{\hat a}}\right)^{n} \qquad \forall \hat a\in\hat{T}_{\mathrm{A,gen}} \quad (25)$$

where $\hat{T}_{\mathrm{A,gen}}$ is the general link set, $\tau_{\hat a}^0$ is the free travel time, parameters $b$, $c$, $n$ are related to road grade and they are obtained by data fitting.

### *B. Multi-period Distribution Network Model*

FCSs are supplied by the electrical buses of DN. The radial DN is represented as $G_{\mathrm{E}}$=[$E_{\mathrm{B}}$, $E_{\mathrm{L}}$], where $E_{\mathrm{B}}$ and $E_{\mathrm{L}}$ denote the bus and line sets, respectively. The power flow equations of DN are constructed based on BFM as follows:

$$P_{ij}(t) - r_{ij}I_{ij}(t) + p_j^{\mathrm{G}}(t) = \sum_{k\in A(j)} P_{jk}(t) + p_j^{\mathrm{L}}(t) \quad \forall j \in E_{\mathrm{B}} \tag{26}$$

$$Q_{ij}(t) - x_{ij}I_{ij}(t) + q_j^{\mathrm{G}}(t) = \sum_{k\in A(j)} Q_{jk}(t) + q_j^{\mathrm{L}}(t) \quad \forall j \in E_B \tag{27}$$

$$U_i(t) - U_j(t) = 2(r_{ij}P_{ij}(t) + x_{ij}Q_{ij}(t)) + {z_{ij}}^2 I_{ij}(t) \ \forall (i,j) \in E_{\mathrm{L}} \tag{28}$$

$$I_{ij}(t)U_i(t) = P_{ij}(t)^2 + Q_{ij}(t)^2 \quad \forall (i,j) \in E_{\mathrm{L}} \tag{29}$$

where $(i, j)$ represents the line from bus $i$ to bus $j$, $A(j)$ is the child bus set of bus $j$, $U$ and $I$ are the squared voltage and current magnitudes, respectively, $r$, $x$ and $z$ are the line resistance, reactance and impedance, respectively, $P$ and $Q$ are the active power and reactive power transmitted on the line, respectively, $p_j^{\mathrm{G}}(t)$ and $p_j^{\mathrm{L}}(t)$ are the active power of the power generation and load demand of bus $j$, respectively, $q_j^{\mathrm{G}}(t)$ and $q_j^{\mathrm{L}}(t)$ are the reactive power of the power generation and load demand of bus $j$, respectively. $p_j^{\mathrm{L}}(t)$ consists of the conventional load $p_j^{\mathrm{lo}}(t)$ and the EV charging load $p_j^{\mathrm{lev}}(t)$, stated as:

$$p_j^{\mathrm{l}}(t) = p_j^{\mathrm{lo}}(t) + p_j^{\mathrm{lev}}(t) \quad = p_j^{\mathrm{lo}}(t) + P_{\mathrm{ev}} \sum_{\hat{a}\in C(j)} x_{\hat{a}}(t) \ \forall j \in E_{\mathrm{B}} \tag{30}$$

where $C(j)$ is the set of charging links supplied by bus $j$.

The operating limits of DN are given as follows:

$$P_{ij}(t) - r_{ij}I_{ij}(t) \ge 0 \quad \forall (i,j) \in E_{\mathrm{L}} \tag{31}$$

$$Q_{ij}(t) - x_{ij}I_{ij}(t) \ge 0 \quad \forall (i,j) \in E_{\mathrm{L}} \tag{32}$$

$$P_{ij}(t)^2 + Q_{ij}(t)^2 \le {S_{\mathrm{lim}}}^2 \quad \forall (i,j) \in E_{\mathrm{L}} \tag{33}$$

$$p_{j,\min}^{\mathrm{G}} \le p_j^{\mathrm{G}}(t) \le p_{j,\max}^{\mathrm{G}} \quad \forall j \in E_{\mathrm{B}} \tag{34}$$

$$-p_{j,\mathrm{ramp}} \le p_j^{\mathrm{G}}(t) - p_j^{\mathrm{G}}(t-1) \le p_{j,\mathrm{ramp}} \quad \forall j \in E_{\mathrm{B}} \tag{35}$$

$$U_{\min} \le U_j(t) \le U_{\max} \quad \forall j \in E_{\mathrm{B}} \tag{36}$$

where constraints (31)-(33) represent the power flow limits, and constraints (34)-(36) represent the power generation and voltage magnitude limits.

The objective of DN operation is to minimize its operating cost, which is stated as:

$$\begin{aligned} \min f_{\mathrm{DN}} &= \sum_t^{T_{\mathrm{DN}}} \sum_{j\in E_{\mathrm{B}}} \left[ a_j p_j^{\mathrm{G}}(t)^2 + b_j p_j^{\mathrm{G}}(t) \right] + \sum_t^{T_{\mathrm{DN}}} \lambda_0(t) \sum_{k\in A(0)} p_{0k}(t) \\ &+ \lambda_{\mathrm{u}} \sum_t^{T_{\mathrm{DN}}} \sum_{j\in E_{\mathrm{B}}} \frac{V_{j0}(t) - \sqrt{U_j(t)}}{V_{j0}(t)} \end{aligned} \tag{37}$$

s.t. (26)-(36)

where the first part is power generation cost, the second part denotes the payment for purchasing electricity from main grids, and the third part is the penalty of bus voltage deviation. $\lambda_0(t)$ is the contract energy price at time $t$, and $\lambda_{\mathrm{u}}$ is the unit cost of voltage deviation. $V_{j0}(t)$ is the voltage of bus $j$ at time $t$ without charging loads.

### C. Coupled Networks Model

Suppose that each EV stops at one FCS and is fully charged during its trip. In TN, the time-varying charging prices and traffic conditions affect the departure time and route choices of EVs, which determines the spatial and temporal variations of charging loads. In DN, the dynamic load demand changes affect the optimal operation of DN, which then impacts LMPs. The interaction between TN and DN reaches an equilibrium state, when no vehicle changes its route or departure time and the optimal operation of DN is achieved. The dynamic equilibrium of the coupled networks is described by the following fixed-point problem:

$$\boldsymbol{x}_{\hat{a}}^*(t) = F_{\mathrm{DTA}}(\boldsymbol{\lambda}^*(t)) = F_{\mathrm{DTA}}(F_{\mathrm{OPF}}(\boldsymbol{x}_{\hat{a}}^*(t))) = F_{\mathrm{CP}}(\boldsymbol{x}_{\hat{a}}^*(t)) \tag{38}$$

where $F_{\mathrm{DTA}}$ and $F_{\mathrm{OPF}}$ represent the DTA and OPF problems, respectively. Note that $\boldsymbol{x}_{\hat{a}}^*(t)$ is the solution of the DTA problem with the given charging price $\boldsymbol{\lambda}^*(t)$, and $\boldsymbol{\lambda}^*(t)$ is the LMP vector obtained from the dual variables of OPF with the given traffic flow $\boldsymbol{x}_{\hat{a}}^*(t)$. For expanded TN with FCSs, $F_{\mathrm{DTA}}$ is presented as:

$$\begin{aligned} &\sum_t^{T_{\mathrm{TN}}} \sum_{\hat{a}\in \hat{T}_{\mathrm{A,ch}}} \left\{ \omega\tau_{\mathrm{ch}} \left[ 1 + \frac{q_{\hat{a}}(t)}{cap_{\hat{a}}} \right] + \lambda_{\hat{a}}^*(t) P_{\mathrm{ev}} \Delta t + T_{\hat{a}}^L \right\} \left[ x_{\hat{a}}(t) \text{-} x_{\hat{a}}^*(t) \right] \\ &+ \sum_t^{T_{\mathrm{TN}}} \sum_{\hat{a}\in \hat{T}_{\mathrm{A,gen}}} \omega\tau_{\hat{a}}^0 \left[ 1 + \left( \frac{x_{\hat{a}}^*(t)}{cap_{\hat{a}}} \right)^{b + c\cdot \left( \frac{x_{\hat{a}}^*(t)}{cap_{\hat{a}}} \right)^n} \right] \left[ x_{\hat{a}}(t) \text{-} x_{\hat{a}}^*(t) \right] \ge 0 \end{aligned} \tag{39}$$

$$\text{s.t.} \quad x_{\hat{a}p}^{rs}(t+1) - x_{\hat{a}p}^{rs}(t) = u_{\hat{a}p}^{rs}(t) - v_{\hat{a}p}^{rs}(t) \quad \forall t, \hat{a}, p, (r,s) \tag{40}$$

$$\sum_{\hat{a}\in A(j)} u_{\hat{a}p}^{rs}(t) = \sum_{\hat{a}\in B(j)} v_{\hat{a}p}^{rs}(t) \quad \forall t, p, (r,s), j \ne r, s \tag{41}$$

$$\sum_p \sum_{\hat{a}\in A(r)} u_{\hat{a}p}^{rs}(t) = \sum_p f_p^{rs}(t) = f^{rs}(t) \quad \forall t, p, (r,s) \tag{42}$$

$$\sum_p \sum_{\hat{a}\in B(s)} v_{\hat{a}p}^{rs}(t) = \sum_p e_p^{rs}(t) = E^{rs}(t+1) \text{-} E^{rs}(t) \quad \forall t, p, (r,s) \tag{43}$$

$$F^{rs} = \sum_{t=1}^{T_{\mathrm{dep}}} f^{rs}(t) \quad \forall (r,s) \tag{44}$$

$$x_{\hat{a}p}^{rs}(t) = \sum_{b\in p'} \left\{ x_{bp}^{rs}[t + \overline{\tau}_{\hat{a}}(t)] - x_{bp}^{rs}(t) \right\} + \left\{ E_p^{rs}[t + \overline{\tau}_{\hat{a}}(t)] - E_p^{rs}(t) \right\} \quad \forall t, \hat{a}, p, (r,s) \tag{45}$$

$$\begin{aligned} &u_{\hat{a}p}^{rs}(t) \ge 0 \quad v_{\hat{a}p}^{rs}(t) \ge 0 \quad x_{\hat{a}p}^{rs}(t) \ge 0 \\ &f_p^{rs}(t) \ge 0 \quad e_p^{rs}(t) \ge 0 \quad \forall t, \hat{a}, p, (r,s) \end{aligned} \tag{46}$$

$$x_{ap}^{rs}(0) = 0 \quad E^{rs}(0) = 0 \quad \forall \hat{a}, p, (r,s) \tag{47}$$

where $T_{\mathrm{dep}}$ is the set of departure time intervals. We discuss the solution's existence and uniqueness of (38) as follows:

*1) Existence*

The solution existence of (38) is demonstrated based on the Brouwer fixed-point theorem.

*Theorem 1 (Brouwer fixed-point theorem):* If $M \subset \mathbf{R}^n$ is convex and compact, then for the continuous self-mapping $F$: $M \to M$, $x \in M$ satisfying $F(x)=x$ exists.

First, we show that the feasible region of the fixed-point problem (38) is convex and compact. Because the charging prices $\boldsymbol{\lambda}(t)$ only affect traveling costs in (39), the feasible region of $F_{\mathrm{CP}}$ is the same as that of $F_{\mathrm{DTA}}$:

$$\Omega_{\mathrm{CP}} = \left\{ \boldsymbol{x}_{\hat{a}}(t) \middle| \text{s.t.(40)-(47)} \right\} \tag{48}$$

The constraints (40)-(47) are linear and all the variables are bounded. Therefore, $\Omega_{\mathrm{CP}}$ is a bounded polytope that is convex and compact.

Second, we make mild assumptions to establish continuous self-mapping of $\Omega_{\mathrm{CP}}$ and prove the solution existence of the fixed-point problem (38).

*Assumption 1:* An EV has at least one feasible path to reach an FCS with adequate charging capacity.

*Assumption 2:* $F_{\text{DTA}}$ is continuous in the charging price $\lambda(t)$.

*Assumption 3:* $\lambda(t)$ is continuous in the charging load demand.

Assumption 1 guarantees that the existence of a DUE state is independent of charging prices, which is usually satisfied in realistic TN. Assumption 2 is satisfied when EV charging power is not very large. Because the travel cost (21) is a linear function of charging prices $\lambda(t)$, a small change of $\lambda(t)$ results in a small travel cost change if $P_{\text{ev}}$ is not very large. Moreover, according to the DUE conditions, the small travel cost results in a small change in traffic flow. Therefore, $F_{\text{DTA}}$ is continuous in $\lambda(t)$. Assumption 3 is satisfied when the charging load accounts for a small part of the total load and the proof is given in [26].

*Proposition 1:* The equilibrium state of coupled networks exists if Assumptions 1-3 are satisfied.

Considering the linear mapping from traffic flow to charging demand in (30), Assumption 3 implies that $F_{\text{OPF}}$ is continuous in $\boldsymbol{x}^*_{\hat{a}}(t)$. Based on Assumption 2, $F_{\text{CP}}(\boldsymbol{x}_{\hat{a}}(t)) = F_{\text{DTA}}(F_{\text{OPF}}(\boldsymbol{x}_{\hat{a}}(t)))$ is continuous in $\boldsymbol{x}^*_{\hat{a}}(t)$. Hence, $F_{\text{CP}}$ is a continuous self-mapping of $\Omega_{\text{CP}}$, and the solution of the fixed-point problem exists.

*2) Uniqueness*

We give the solution uniqueness condition based on the Contraction mapping theorem.

*Theorem 2 (Contraction mapping theorem)*: If the continuous self-mapping is a contraction mapping, the solution to the fixed-point problem is unique.

Based on the contraction mapping definition, in the metric space ($\Omega_{\text{CP}}$, $\rho$), if there is a constant $k \in (0, 1)$, for any $\boldsymbol{x}_{\hat{a}1}(t)$, $\boldsymbol{x}_{\hat{a}2}(t) \in \Omega_{\text{CP}}$:

$$\frac{\rho(F_{\text{CP}}(\boldsymbol{x}_{\hat{a}1}(t)), F_{\text{CP}}(\boldsymbol{x}_{\hat{a}2}(t)))}{\rho(\boldsymbol{x}_{\hat{a}1}(t), \boldsymbol{x}_{\hat{a}2}(t))} \le k \tag{49}$$

then $F_{\text{CP}}$ is a contraction mapping. Combined with Lagrange mean value theorem, constraint (49) is further transformed into:

$$\rho(\nabla F_{\text{CP}}(\boldsymbol{x}_{\hat{a}}(t))) \le 1 \tag{50}$$

where $\rho(\nabla F_{\text{CP}}(\boldsymbol{x}_{\hat{a}}(t)))$ is the spectral radius of the Jacobian matrix of $F_{\text{CP}}$. As long as constraint (50) is satisfied, $F_{\text{CP}}$ is likely to be contractive and the fixed-point problem has a unique solution.

## IV. Solution Method

In this section, an efficient algorithm is designed to solve the fixed-point problem of coupled TN and DN, and then an MCS technique is used to evaluate the influence of variable travel demands on the operation of TN and DN.

### *A. Solution Method of Coupled Networks*

The solution method of coupled networks relies on the reliable and efficient solution of DTA and OPF. However, there exist nonlinearity and non-convexity in the two problems. Hence, the linearization and convex relaxation techniques are used to improve computational efficiency.

*1) Linearization of the DTA model*

There are three sources of nonlinearity in the DTA model. First, the dynamic queue flow $q_{\hat{a}}(t)$ in (39) is described by the following linear constraints:

$$q_{\hat{a}}(t) \ge 0 \qquad \forall \hat{a} \in \hat{T}_{\text{A,ch}} \tag{51.1}$$

$$q_{\hat{a}}(t) \ge x_{\hat{a}}(t) - cap_{\hat{a}} \qquad \forall \hat{a} \in \hat{T}_{\text{A,ch}} \tag{51.2}$$

$$q_{\hat{a}}(t) \le M\left(1 - u_{\hat{a},1}(t)\right) \qquad \forall \hat{a} \in \hat{T}_{\text{A,ch}} \tag{51.3}$$

$$q_{\hat{a}}(t) \le \left(x_{\hat{a}}(t) - cap_{\hat{a}}\right) + M\left(1 - u_{\hat{a},2}(t)\right) \qquad \forall \hat{a} \in \hat{T}_{\text{A,ch}} \tag{51.4}$$

$$u_{\hat{a},1}(t) + u_{\hat{a},2}(t) \ge 1 \quad u_{\hat{a},1}(t), u_{\hat{a},2}(t) \in \{0,1\} \qquad \forall \hat{a} \in \hat{T}_{\text{A,ch}} \tag{51.5}$$

where $u_{\hat{a},1}(t)$ and $u_{\hat{a},2}(t)$ are auxiliary binary variables for each $q_{\hat{a}}(t)$, and $M$ is a large positive constant.

Second, the travel cost function of general links in (39) is approximated by the piecewise linear functions:

$$\begin{gathered} x_{\hat{a}}(t) = \sum_m \alpha^m_{\hat{a}}(t) x^m_{\hat{a}} \quad \sum_m \alpha^m_{\hat{a}}(t) = 1 \quad c_{\hat{a}}(t) = \sum_m c_{\hat{a}}(x^m_{\hat{a}}) \alpha^m_{\hat{a}}(t) \\ \left\{\alpha^m_{\hat{a}}(t), \forall m\right\} \in \text{SOS2} \quad \forall \hat{a} \in \hat{T}_{\text{A,gen}} \end{gathered} \tag{52}$$

where $x^m_{\hat{a}}$ is the breakpoint of each sub-interval, and $\{\alpha^m_{\hat{a}}(t), \forall m\}$ is a vector in the special ordered set of type 2 (SOS2). SOS2 denotes an ordered set of non-negative variables, of which at most two adjacent variables are non-zero [27].

Third, for the nonlinear propagation constraint (45), according to **Algorithm 1**, the estimated travel time of link $a$ at time $t$ is equal to fixed $\bar{\tau}^n_{\hat{a}}(t)$ in the $n$th loop; thus, constraint (45) is linear in the solution procedure. Finally, the DTA problem is converted into mixed-integer linear programming (MILP).

*2) Convex relaxation of OPF model*

The non-convexity of the OPF problem comes from the quadratic constraint (29). Based on the second-order cone (SOC) relaxation technique [28], it is replaced by the SOC inequality:

$$\left\| \begin{matrix} 2P_{ij}(t) \\ 2Q_{ij}(t) \\ I_{ij}(t) - U_i(t) \end{matrix} \right\|_2 \le I_{ij}(t) + U_i(t) \qquad \forall (i,j) \in E_{\text{L}} \tag{53}$$

The SOC relaxation is exact for radial networks under mild conditions that are satisfied in our model. Next, the quadric term $a_j p^{\text{G}}_j(t)^2$ and square term $\sqrt{U_j(t)}$ in (37) are replaced by auxiliary variables $s_j(t)$ and $V_j(t)$, and the following SOC inequalities are considered:

$$\left\| \begin{matrix} 2\sqrt{a_j}\, p^{\text{G}}_j(t) \\ s_j(t) - 1 \end{matrix} \right\|_2 \le s_j(t) + 1 \qquad \forall j \in E_{\text{L}} \tag{54}$$

$$\left\| \begin{matrix} 2V_j(t) \\ U_j(t) - 1 \end{matrix} \right\|_2 \le U_j(t) + 1 \qquad \forall j \in E_{\text{L}} \tag{55}$$

Therefore, the OPF problem is transformed into a second-order cone programming (SOCP) problem that is more tractable for commercial solvers. The equilibrium state of TN and DN is the solution of the fixed-point problem $F_{\text{CP}}$ and is obtained by the decomposition method as follows:

**Algorithm 2:** Decomposition method

**1 Input:** $i$=1, $G_E$, $\hat{G}_T$, $F^{rs}$, initial $\lambda(t)^i$, initial $\boldsymbol{x}_{\hat{a}}(t)^i$

**2 repeat**

**3** Solve the OPF problem (37), and get dual variables $\lambda(t)^{i+1}$

Let $\Delta DN = \dfrac{\left\| \lambda(t)^{i} - \lambda(t)^{i+1} \right\|_2}{\left\| \lambda(t)^{i} \right\|_2}$

**4** Solve the DTA problem (39)-(47) by **Algorithm 1**, and get traffic flow $\boldsymbol{x}_{\hat{a}}(t)^{i+1}$

Let $\Delta TN = \dfrac{\left\| \boldsymbol{x}_{\hat{a}}(t)^{i} - \boldsymbol{x}_{\hat{a}}(t)^{i+1} \right\|_2}{\left\| \boldsymbol{x}_{\hat{a}}(t)^{i} \right\|_2}$

**5** Loop index: $i$=$i$+1

**6** **until** $\Delta DN + \Delta TN < \varepsilon$

**7** **Output:** $\lambda(t)$, $\boldsymbol{x}_{\hat{a}}(t)^{i}$

### B. Monte Carlo simulation of coupled networks

A random variable $q_{rs}$ is used to describe the uncertain travel demand between O-D pair ($r$, $s$) and it is assumed to follow the Normal distribution. Besides, travel demands are spatially correlated because of traveling habits. Here, the Pearson correlation is used to describe the correlations of the travel demands between different O-D pairs, and it can be obtained from the realistic condition. The covariance of the travel demands between O-D pairs ($r_1$, $s_1$) and ($r_2$, $s_2$) is:

$$\mathrm{cov}(q_{rs1}, q_{rs2}) = \rho(q_{rs1}, q_{rs2})\sigma_{rs1}\sigma_{rs2} \tag{56}$$

where $q_{rs1}$ and $q_{rs2}$ represent the uncertain travel demands between the O-D pairs ($r_1$, $s_1$) and ($r_2$, $s_2$), respectively. $\rho(q_{rs1}, q_{rs2})$ is the Pearson correlation coefficient of the random variables $q_{rs1}$ and $q_{rs2}$. $\sigma_{rs1}$ and $\sigma_{rs2}$ are the standard deviations of the travel demands.

Based on the stochastic travel demand and the fixed-point problem, the stochastic equilibrium states of coupled networks are obtained by MCS, as shown in Fig. 3.

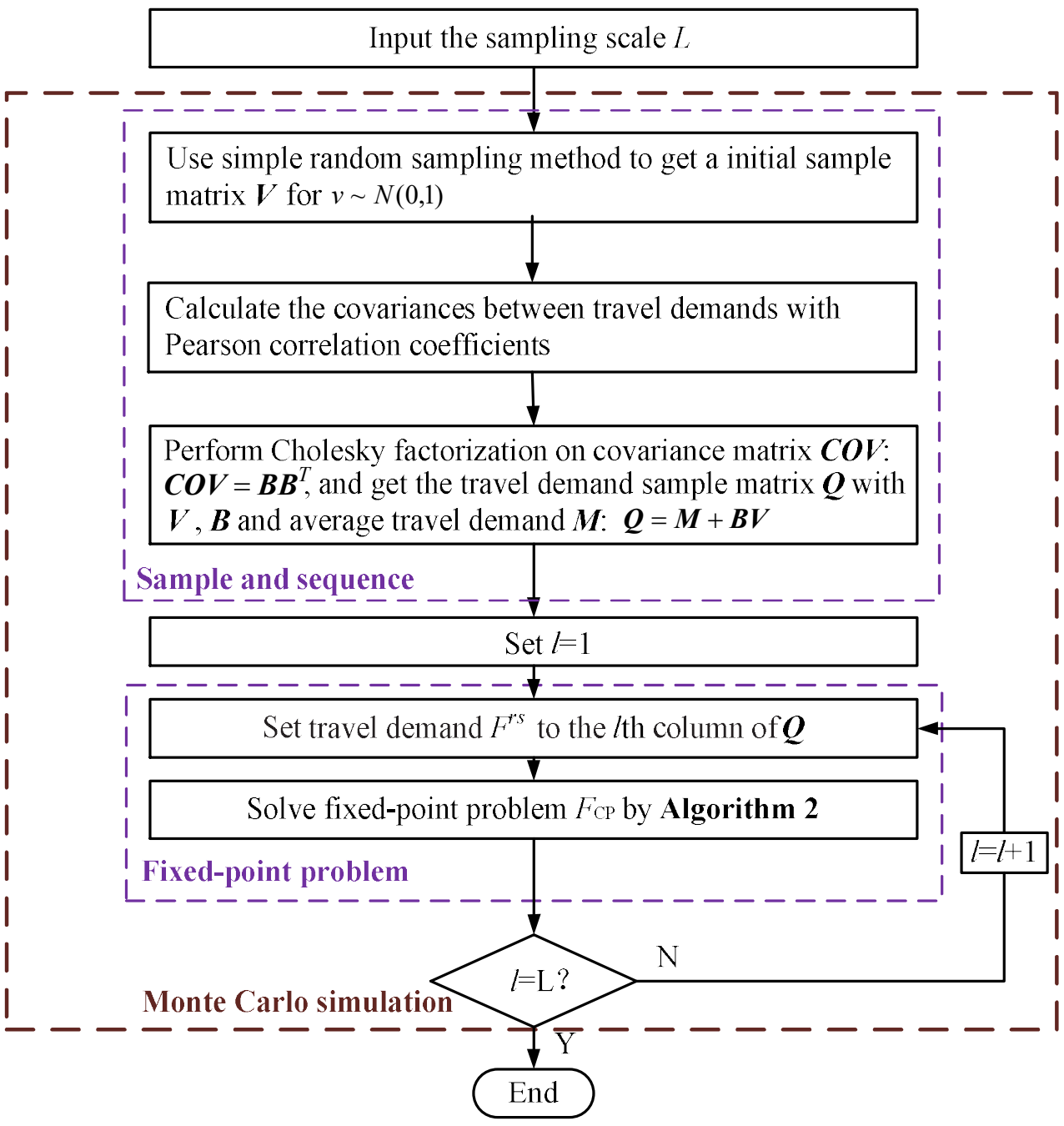

Fig. 3. Flowchart of Monte Carlo simulation.

## V. Case Studies

In this section, numerical experiments are carried out to verify the effectiveness, correctness and extensibility of the proposed coupled networks model.

### A. Effectiveness of the designed coupled networks model

Fig. 4 shows the topologies of a ring form TN and a radial DN. The basic parameters of TN and DN are given in Tables I and II, and the deterministic O-D demands are shown in Table III. The base power is 100 MVA. The base traffic flow is designed as 30 vehicles/h, the travel time monetary value $\omega$ is 10 \$/h and the load fluctuation monetary value $\alpha$ is 1 \$/p.u.. The time duration is one hour, which is divided into 12 and 6 intervals in DTA and OPF. The convergence tolerances are set as $\varepsilon$=0.05, $\varepsilon_1$=0.01, $\varepsilon_2$=0.01. We design five cases in Table IV to compare different coupled networks models and illustrate the advantages of our work.

TABLE I
LINK CAPACITY AND FREE TRAVEL TIME/CHARGING TIME

| Link type | Type 1 | Type 2 | Type 3 | Type 4 | Charging |
|---|---|---|---|---|---|
| $cap_a$ (p.u.) | 100 | 100 | 60 | 80 | 20 |
| $\tau^0_a/\tau_{ch}$ (min) | 8 | 5 | 7 | 5 | 10 |

TABLE II
PARAMETERS IN DN

| $a$ (\$/MW$^2$h) | $b$ (\$/MWh) | $\lambda_0$ (\$/MWh) | $\lambda_u$ (\$/p.u.) |
|---|---|---|---|
| 0.3 | 130 | 140 | 200 |
| $U_{max}$ (p.u.$^2$) | $U_{min}$ (p.u.$^2$) | $p^G_{max}/p^G_{min}$ (p.u.) | $q^G_{max}/q^G_{min}$ (p.u.) |
| 1.1025 | 0.7744 | 0/1.5 | -1/1 |
| $S_{lim}^2$ (p.u.$^2$) | $p^{lo}/q^L$ (p.u.) | $P_{ev}$ (kW) | $p_{ramp}/q_{ramp}$ (p.u.) |
| 0.01 | 0.1 | 150 | 0.5/0.4 |

TABLE III
DETERMINISTIC O-D DEMANDS (P.U.)

| O-D | $F^{rs}_{gas}$ | $F^{rs}_{ev}$ | O-D | $F^{rs}_{gas}$ | $F^{rs}_{ev}$ |
|---|---|---|---|---|---|
| 1-6 | 30 | 10 | 3-12 | 30 | 10 |
| 1-10 | 60 | 5 | 3-11 | 50 | 5 |
| 1-12 | 40 | 10 | 4-9 | 40 | 10 |
| 1-11 | 40 | 5 | 4-10 | 50 | 5 |
| 3-6 | 50 | 10 | 4-12 | 40 | 10 |
| 3-10 | 40 | 10 | | | |

TABLE IV
CONTRASTIVE CASES

| | | Case 1 | Case 2 | Case 3 | Case 4 | Case 5 |
|---|---|---|---|---|---|---|
| TN | Traffic flow model | DTA | STA | DTA | DTA | DTA |
| | Travel choice | departure time/route choice | route choice | route choice | departure time/route choice | departure time/route choice |
| | Travel cost | PWL | PWL | PWL | PWL | nonlinear |
| | Load fluctuation toll | √ | × | √ | × | √ |
| DN | Power flow model | multi-period OPF | OPF | multi-period OPF | multi-period OPF | multi-period OPF with (29) |
| | Bus voltage deviation | √ | √ | √ | × | √ |

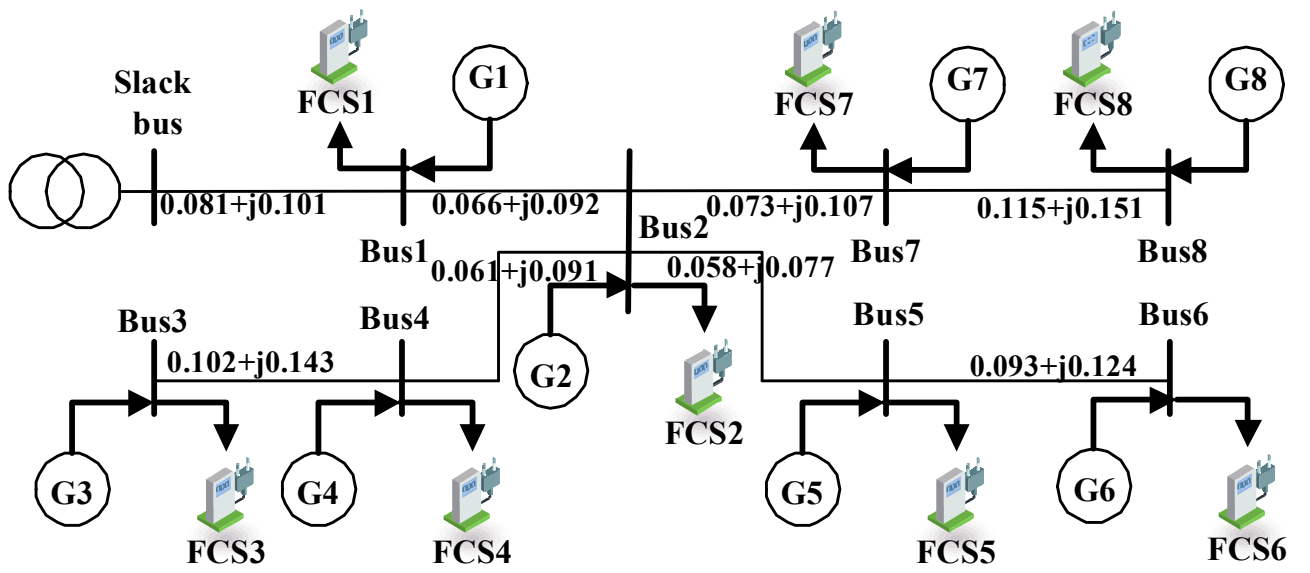

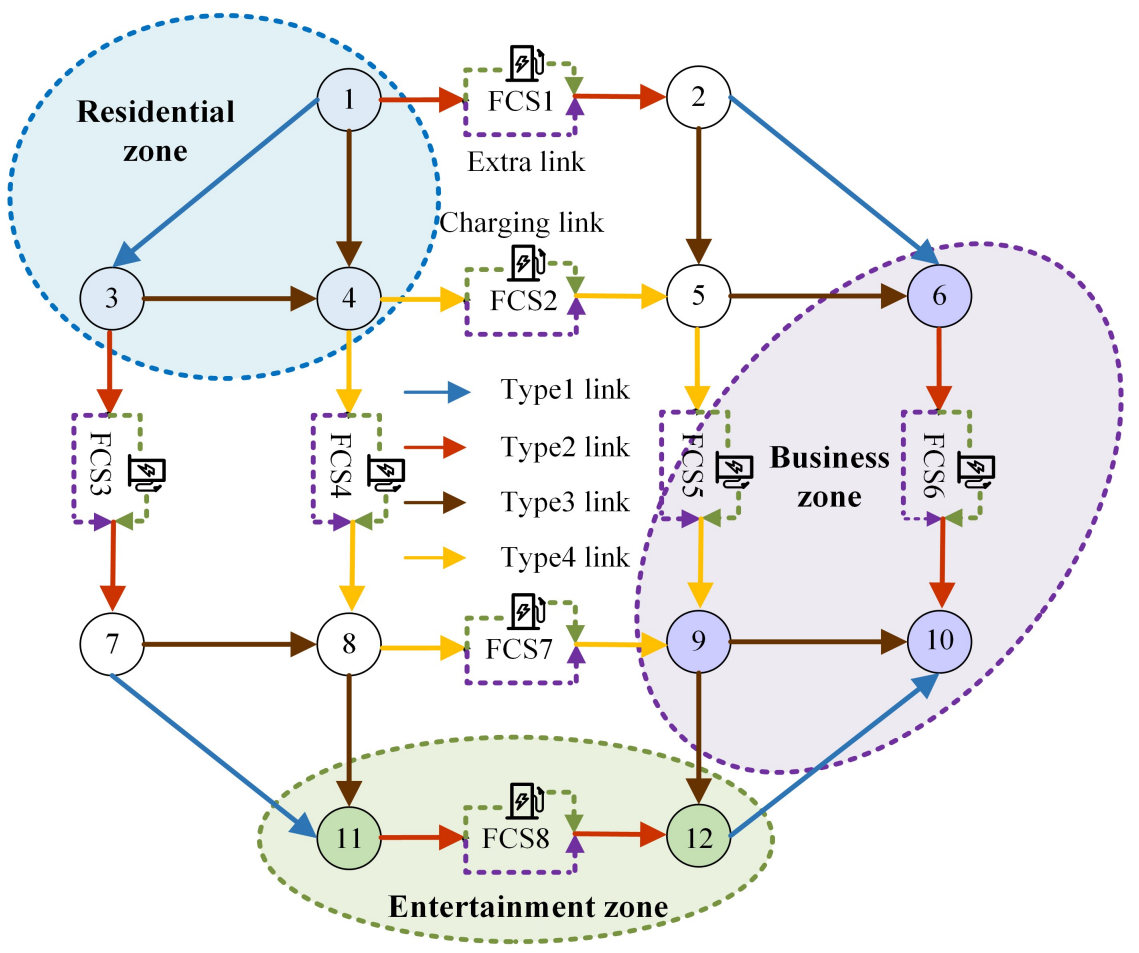

Fig. 4. Topologies of TN and DN.

*1) Effectiveness of the DTA model with both departure time and route choices:*

The effectiveness of utilizing the DTA with both departure time and route choices in our coupled networks model is verified by the comparison of Cases 1, 2 and 3.

In Fig. 5(a), the average charging demands in Case 2 are much larger than those in Case 1. This is because the STA model neglects that different EVs are charged at different periods, and the charging load at each FCS remains constant in the simulation duration, which leads to the overestimation of charging demand. On the contrary, the charging loads in the DTA model are distributed in different periods by capturing the dynamic flow propagation, which makes the average charging demands smaller. Therefore, in Fig. 5(b), the average electricity prices in Case 1 are always smaller than those in Case 2 with the more accurate distributions of charging load in the time domain, and it is beneficial for the economic operation of the two systems.

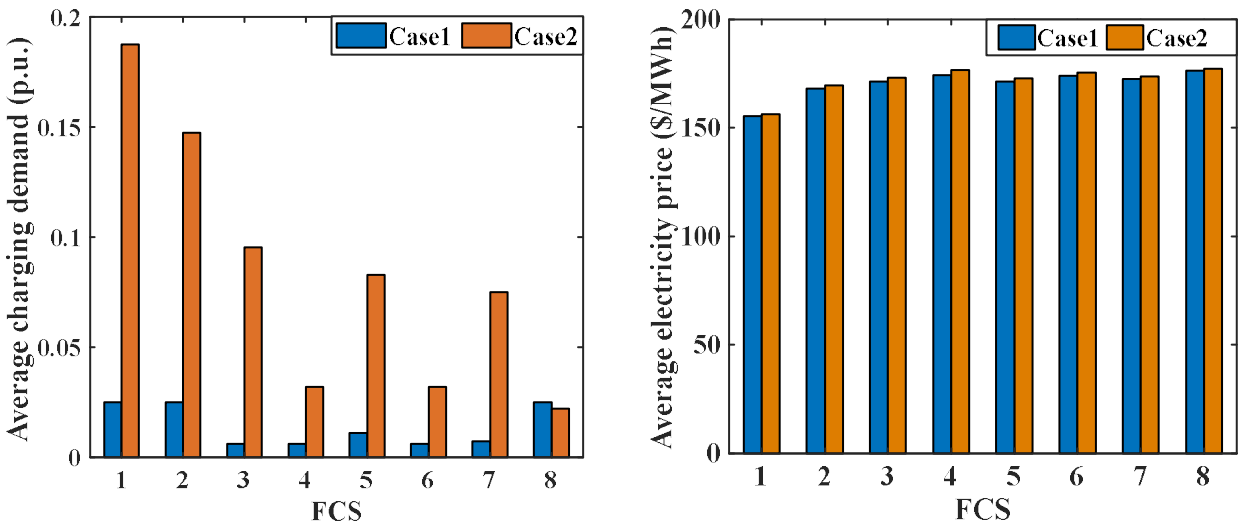

(a) Average charging demands (b) Average electricity prices

Fig. 5. Average charging demands and electricity prices in Cases 1 and 2.

Taking the O-D pair (1, 6) as an example, Tables V and VI present the travel costs and departure rates for EVs and GVs in Cases 1 and 3. There are four candidate paths: 1→2→6, 1→2→5→6, 1→4→5→6, 1→3→4→5→6 in two cases, and three candidate departure time intervals: 1, 2, 3 in Case 1. Compared with the results in Case 1, the minimal GV and EV travel costs in Case 3 are increased by 10.2% and 7.4%, respectively. This is because Case 3 does not consider the departure time choice and the traffic congestion occurs with all the travelers departing at the same time. Thus, the DTA model with both departure time and route choices improves the traffic efficiency of TN. Moreover, Fig. 6 gives the distributions of charging loads in Cases 1 and 3. The charging loads are diversely distributed and heavy loads are partially diminished in Case 1 with the departure time choice. For example, the maximum charging load in Case 1 is 5.63 MW, which is smaller than the 7.50 MW in Case 3. This is beneficial for DN, because the heavy charging loads may cause large voltage deviations and affect the system's safety.

Therefore, the designed coupled networks model with the DTA considering both departure time and route choices not only describes traffic flows more accurately, but also benefits the operation of TN and DN.

TABLE V
TRAVEL COSTS AND DEPARTURE RATES OF O-D PAIR (1, 6) IN CASE1 (P.U.)

| Path | $c_{gv}(1)$ | $c_{gv}(2)$ | $c_{gv}(3)$ | $c_{ev}(1)$ | $c_{ev}(2)$ | $c_{ev}(3)$ |
|---|---|---|---|---|---|---|
| 1→2→6 | 2.186 | 2.186 | 2.186 | 7.685 | 7.701 | 7.779 |
| 1→2→5→6 | 3.232 | 3.222 | 3.213 | 8.729 | 8.690 | 8.789 |
| 1→4→5→6 | 3.282 | 3.278 | 3.275 | 9.293 | 9.011 | 9.781 |
| 1→3→4→5→6 | 4.591 | 4.502 | 4.513 | 10.933 | 10.369 | 10.450 |
| Path | $f_{gv}(1)$ | $f_{gv}(2)$ | $f_{gv}(3)$ | $f_{ev}(1)$ | $f_{ev}(2)$ | $f_{ev}(3)$ |
| 1→2→6 | 11 | 14 | 5 | 10 | 0 | 0 |
| 1→2→5→6 | 0 | 0 | 0 | 0 | 0 | 0 |
| 1→4→5→6 | 0 | 0 | 0 | 0 | 0 | 0 |
| 1→3→4→5→6 | 0 | 0 | 0 | 0 | 0 | 0 |

TABLE VI
TRAVEL COSTS AND DEPARTURE RATES OF O-D PAIR (1, 6) IN CASE3 (P.U.)

| Path | $c_{gv}$ | $f_{gv}$ | $c_{ev}$ | $f_{ev}$ |
|---|---|---|---|---|
| 1→2→6 | 2.409 | 30 | 8.250 | 10 |
| 1→2→5→6 | 3.423 | 0 | 9.341 | 0 |
| 1→4→5→6 | 3.459 | 0 | 9.345 | 0 |
| 1→3→4→5→6 | 4.512 | 0 | 10.297 | 0 |

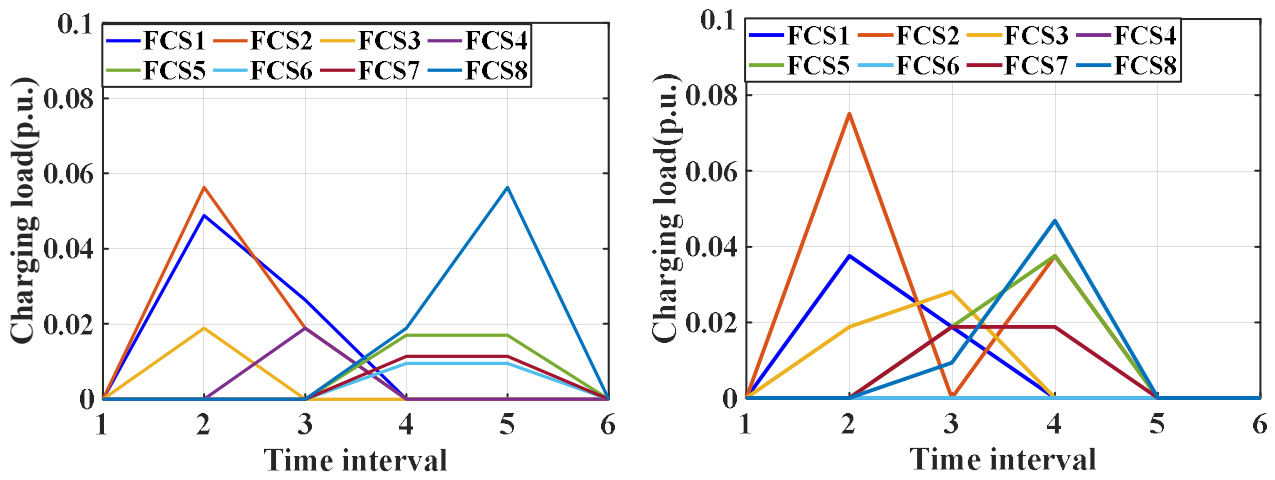

(a) Case 1 (b) Case 3

Fig. 6. Distributions of charging loads in Cases 1 and 3.

*2) Effectiveness of considering bus voltage deviation and charging load fluctuation:*

Fig. 7 presents the bus voltage and its deviation in Cases 1 and 4. Compared with Fig. 7(b), the maximum voltage deviation is decreased by 43.2% in Fig. 7(a). It is because when the bus voltage deviation index is considered in the objective function of OPF, the large voltage deviation increases the LMP of the bus, and EVs will choose other FCSs with lower electricity prices, thus reducing the charging load and voltage deviation at this bus. In addition, by considering the charging load fluctuation toll in DTA, the voltage variations among different time intervals become smaller, as shown in Fig. 7(c). For example, the standard deviation of voltages at bus 3 is 0.009 in Case 1, which is much smaller than the 0.016 in Case 4. The charging load fluctuation toll directly influences the travel choices of EVs and makes the charging load distributions smoother, which further reduces the voltage fluctuation in DN.

Therefore, by considering the bus voltage deviation index and charging load fluctuation toll in the designed coupled networks model, the adverse influence of the interaction, especially the voltage quality problem caused by remarkable charging load demands has been diminished. This is essential for the safe operation of DN with a large amount of EVs.

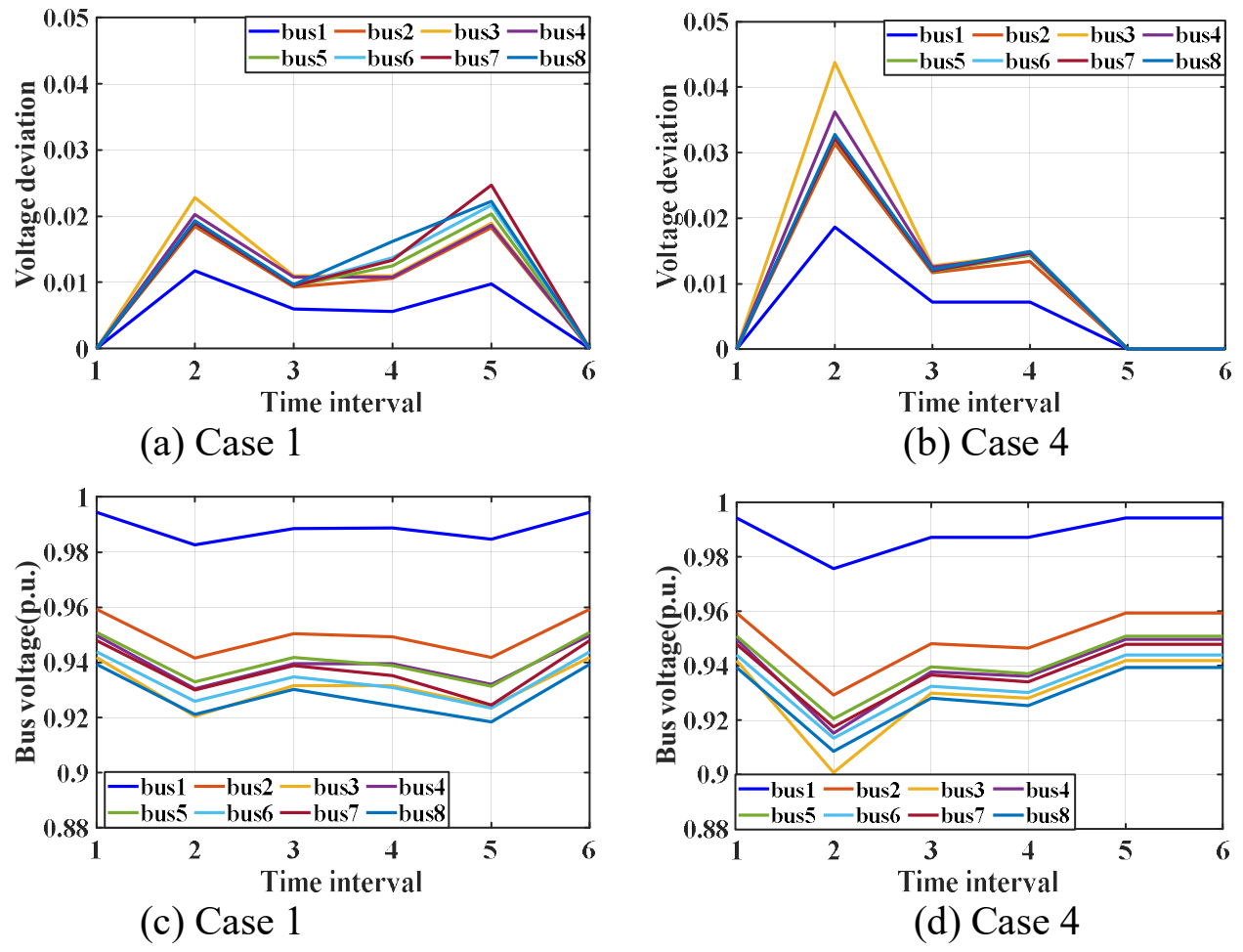

(a) Case 1 (b) Case 4

(c) Case 1 (d) Case 4

Fig. 7. Bus voltages and the deviations in Cases 1 and 4.

*3) Computational efficiency and accuracy:*

The computation times of Cases 1 and 5 are 252s and 7851s, respectively. Hence, the piecewise linearization (PWL) approximation of travel costs in DTA and the convex relaxation in OPF significantly improve computational efficiency.

The total deviation between $U_i(t)I_{ij}(t)$ and $P_{ij}(t)^2 + Q_{ij}(t)^2$ in Case 1 is 1.537e-7, which verifies the exactness of convex relaxation. Moreover, the approximation errors of traffic flow and electricity price between Cases 1 and 5 are 3.813% and 1.032%, respectively, which indicates that the accuracy of our solution method is maintained as well.

## *B. Coupled Networks interaction analysis*

In this subsection, we use the developed coupled networks model to analyze the interaction between TN and DN. The test systems are the same as part A. First, the travel demands of TN are designed as deterministic inputs, and the following two cases are used to evaluate the role of price signals in coordinating TN and DN.

Case 6: There is no information exchange between TN and DN. EVs do not know charging prices before arriving at FCSs and select the departure time and route based on travel time.

Case 7: There is a price information exchange between TN and DN. EVs obtain the price information in real time and consider the charging cost in their total travel cost.

Fig. 8 gives the average electricity prices and the differences of charging EVs and passing GVs at each FCS in two cases. Compared with the results in Case 6, the EV numbers are decreased at FCSs with higher prices and are increased at those with lower prices in Case 7. This is because EVs reduce their charging costs when real-time electricity prices are available. The changes of EV charging choices further affect the traffic flow distribution. As shown in Fig. 8, the changes of GVs are contrary to those of EVs at each FCS to avoid traffic congestion. Moreover, the changes of traffic flow distributions have influenced DN simultaneously. In Case 7, the electricity prices at FCSs are reduced with the decrease of the number of charging EVs. The above alternating process shows the interaction between TN and DN, and the equilibrium states of coupled networks are achieved eventually. Fig. 9 presents the bus voltage distributions in Cases 6 and 7. Because the bus voltage deviation is considered as a penalty in the DN operating cost, the voltage deviations are much smaller when the price information is exchanged between the two systems.

Finally, the stochastic analyses are performed to evaluate the interaction between TN and DN in uncertain conditions. The uncertain travel demands follow Normal distributions with means given in Table III and coefficients of variation equal to 0.1. The sample size is set as 200, and correlation coefficients between different O-D demands are given in Table VII.

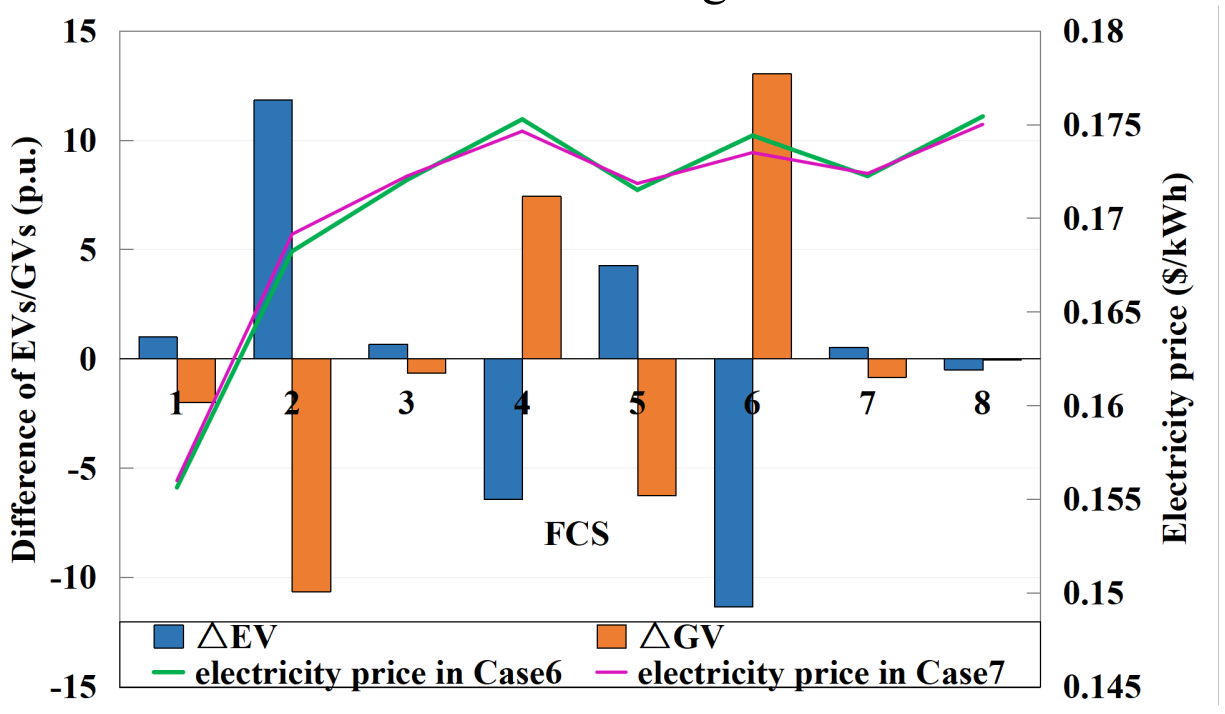

Fig. 8. Average electricity prices of Cases 6 and the differences of EVs/ GVs at each FCS between Cases 6 and 7.

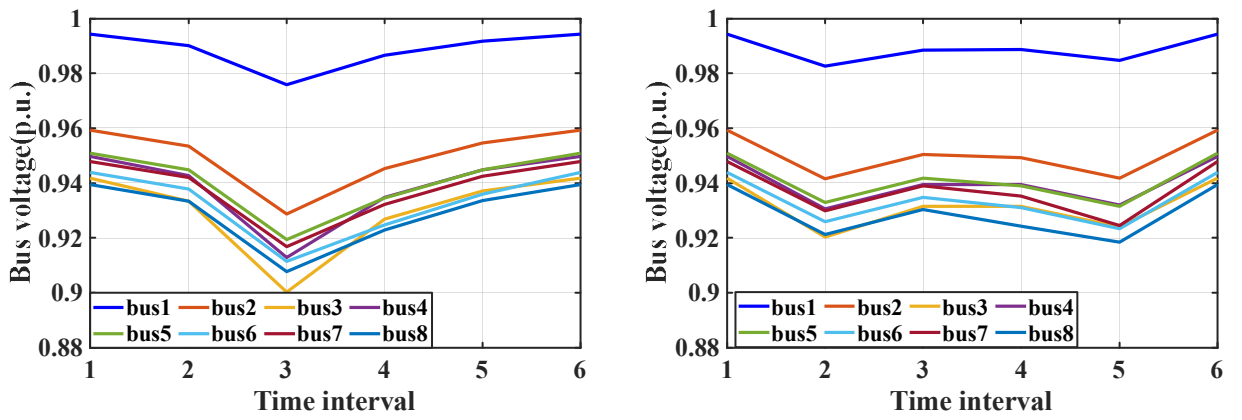

(a) Case 6 (b) Case 7

Fig. 9. Bus voltage distribution in Cases 6 and 7.

TABLE VII

CORRELATION COEFFICIENTS BETWEEN O-D DEMANDS

| | Destinations in the same zones | Destinations in different zones |
|---|---|---|
| Same origin | 0.3 | -0.1 |
| Different origins | 0.15 | -0.05 |

Fig. 10 gives the probability density distributions of average electricity prices and EV charging flows at each FCS in Cases 6 and 7. Compared with Fig. 10(c), the charging EV flows at FCSs 1 and 2 are much larger in Fig. 10(d). This is because the price information is exchanged between TN and DN in Case 7 and more EVs choose to charge at the two FCSs because of smaller prices (see Fig. 10(a) and (b)). Similarly, the charging flows at FCSs 4 and 8 are decreased in Case 7, which indicates the influences of coupled networks interaction.

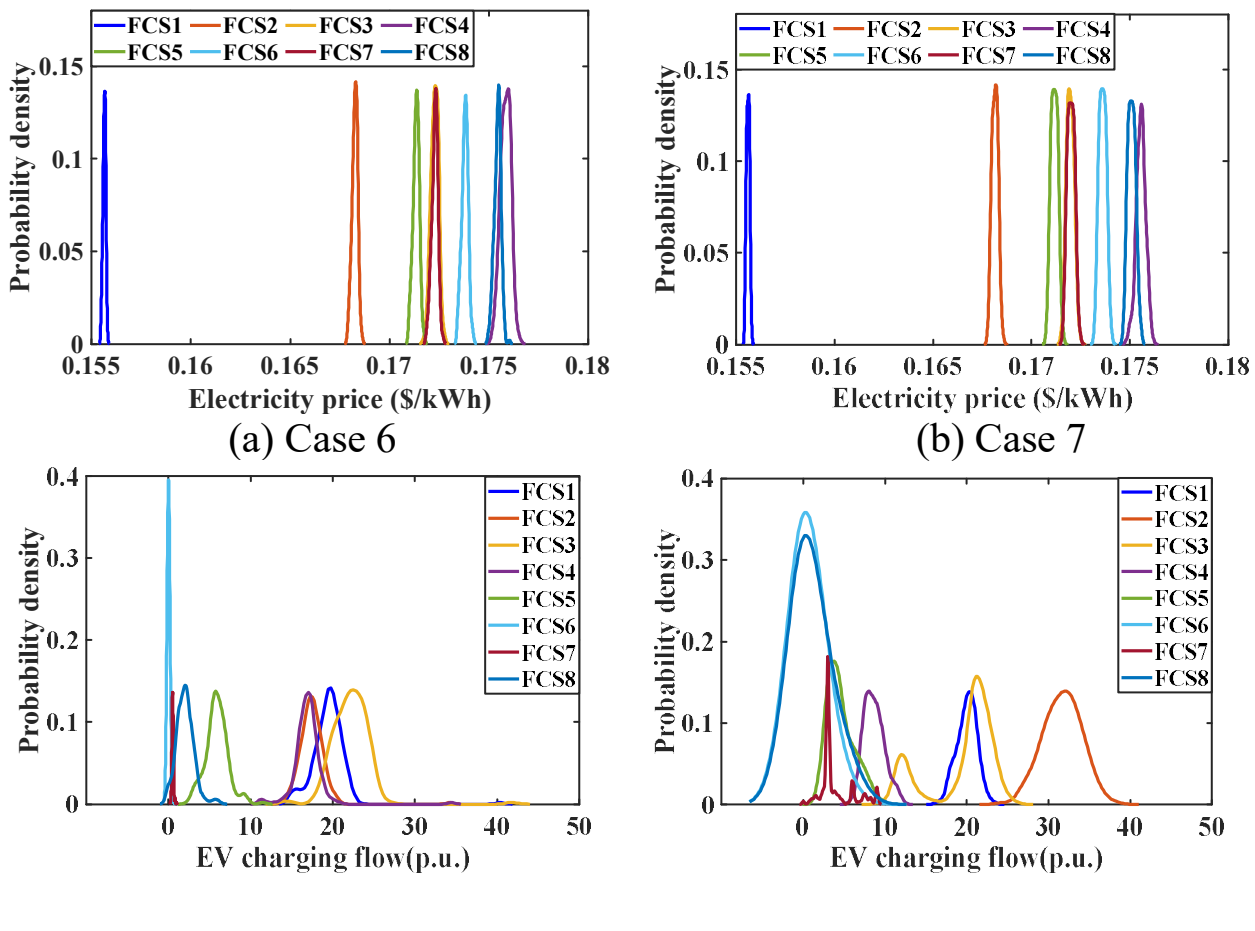

(a) Case 6 (b) Case 7

(c) Case 6 (d) Case 7

Fig. 10. Probability densities of average electricity prices and total charging loads in Cases 6 and 7.

On the other hand, the interaction between two networks increases the operational risks of both systems. Fig. 11 gives the standard deviations of EV charging flows in Cases 6 and 7, and the dispersion degree is significantly larger in Case 7, which can also be seen from the comparison of Fig. 10(c) and (d). This is because when the interaction between TN and DN is considered, the uncertain charging loads caused by the uncertain travel demands affect the operation of DN, and make the electricity prices changeable. Furtherly, the uncertain electricity prices affect the travel choices of EVs and the traffic flow distributions, leading to more variable system states. In summary, the bi-directional influences between TN and DN increase the uncertainties of coupled networks, and should be considered in evaluating the stochastic states of both systems.

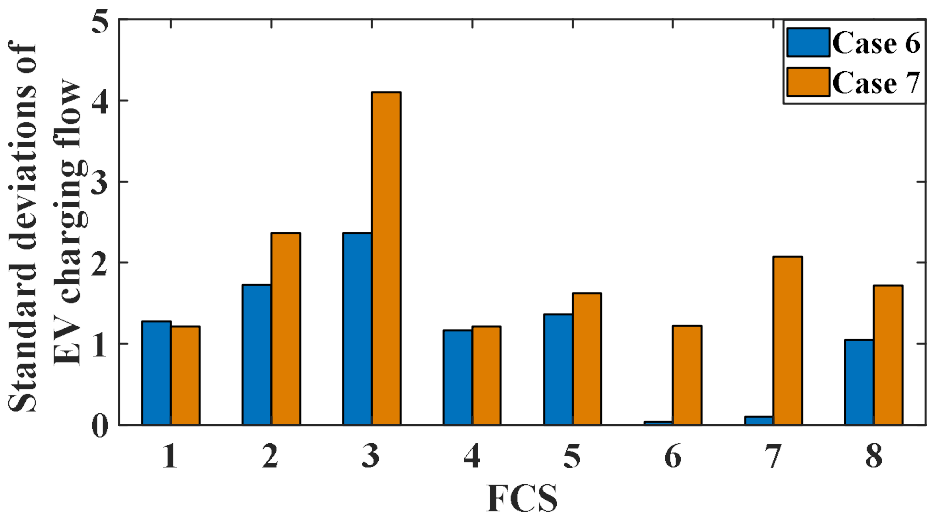

Fig. 11. Standard deviations of EV charging flows in Cases 6 and 7.

## C. Correctness and extensibility of the designed model

In this subsection, we combine a realistic transportation system of Sioux Falls [29] with the US PG&E 69-bus DN [30] to examine the correctness and extensibility of the designed model. Fig. 12 shows the topologies of two systems. The base voltage is 12.66 kV and the base power is 10 MVA. The lower and upper limits of bus voltages are 0.95 p.u. and 1.05 p.u., respectively. $S_{\text{lim}}^2$ is set as 0.36 p.u.$^2$. The base traffic flow is 30 vehicles/h. The simulation duration is 1.5 h, which is divided into 18 and 9 intervals in DTA and OPF. Considering the growth trend of vehicles in the US and the proportion of EVs predicted by the Energy Information Administration (EIA), the real-world travel demands of six O-D pairs in [29] have been expanded (see Table VIII).

TABLE VIII
PREDICTED O-D DEMANDS IN 2050 (P.U.)

| O-D | $F^{rs}_{\text{gas}}$ | $F^{rs}_{\text{ev}}$ | O-D | $F^{rs}_{\text{gas}}$ | $F^{rs}_{\text{ev}}$ |
|---|---|---|---|---|---|
| 1-10 | 40.9 | 12.7 | 10-20 | 78.6 | 24.4 |
| 4-17 | 15.7 | 4.9 | 12-20 | 12.6 | 3.9 |
| 5-10 | 31.4 | 9.7 | 23-20 | 22.0 | 6.8 |

Taking O-D pair (23, 20) as an example, there are three paths for GVs (23→24→21→20, 23→22→20, 23→22→21→20) and three paths for EVs (23→FCS6→22→20, 23→22→FCS5→20, 23→FCS6→21→20) to choose, and the number of candidate departure time intervals is set as three. When the coupled networks reach the equilibrium state in our designed model, the travel cost and departure rate for each path at each departure time are shown in Table IX. It can be seen that the travel costs of the chosen paths at the chosen departure time for both EVs and GVs are equal and minimal, and no travelers choose those travel schemes with larger travel costs. It satisfies the DUE conditions in equation (11) and conforms to the selfishness of individual travelers in reality. Thus, the correctness of the DUE conditions with both departure time and route choices in our designed model is verified.

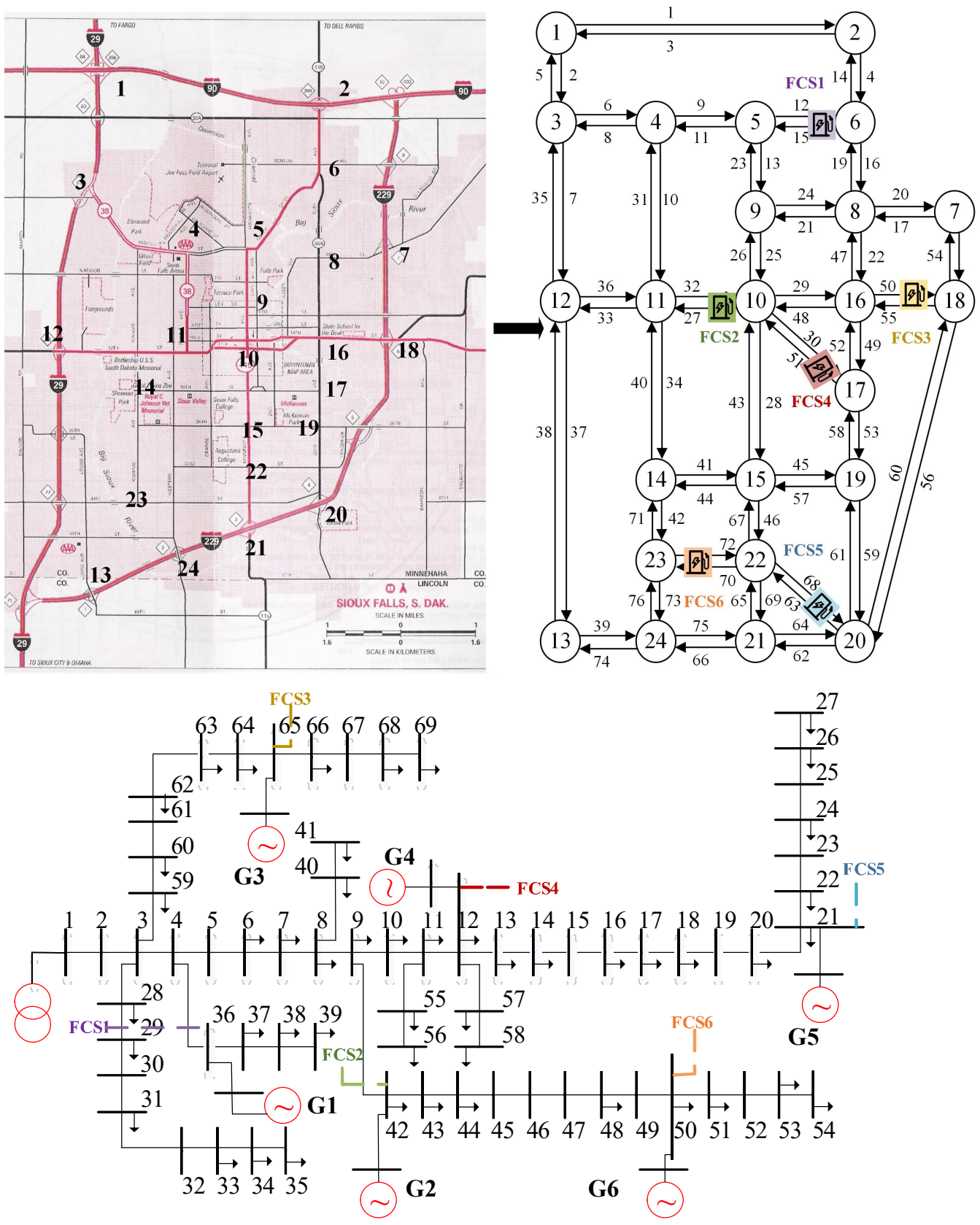

Fig. 12. Topologies of Sioux Falls TN and US PG&E 69-bus DN.

TABLE IX
TRAVEL COSTS AND DEPARTURE RATES OF O-D PAIR (23, 20) (P.U.)

| Path | $c_{\text{gv}}(1)$ | $c_{\text{gv}}(2)$ | $c_{\text{gv}}(3)$ | $c_{\text{ev}}(1)$ | $c_{\text{ev}}(2)$ | $c_{\text{ev}}(3)$ |
|---|---|---|---|---|---|---|
| 1 | 3.50 | 3.50 | 3.50 | 7.20 | 7.20 | 7.29 |
| 2 | 3.50 | 3.50 | 3.50 | 7.06 | 7.06 | 7.06 |
| 3 | 2.33 | 2.33 | 2.33 | 8.22 | 8.23 | 8.23 |
| Path | $f_{\text{gv}}(1)$ | $f_{\text{gv}}(2)$ | $f_{\text{gv}}(3)$ | $f_{\text{ev}}(1)$ | $f_{\text{ev}}(2)$ | $f_{\text{ev}}(3)$ |
| 1 | 0 | 0 | 0 | 0 | 0 | 0 |
| 2 | 0 | 0 | 0 | 2.11 | 2.36 | 2.36 |
| 3 | 11.77 | 7.59 | 2.65 | 0 | 0 | 0 |

Fig.13 shows the bus voltages and the squared apparent power of the line under the equilibrium state in our model. Different colors represent different buses or lines. In Fig. 13(a), the bus voltages are within 0.95 p.u.~1.05 p.u., which satisfies the voltage magnitude limits, and the squared apparent power of the line in Fig. 13(b) is also no more than 0.36 p.u.$^2$, which satisfies the power flow limit. Thus, the correctness of our coupled networks model in DN operation is verified.

The calculation time for this large system is 338s, and the solution procedure converges in three iterations. The proposed model has good extensibility and can be utilized to analyze the dynamic interaction between two systems and realize the economic and safe operation of realistic coupled networks.

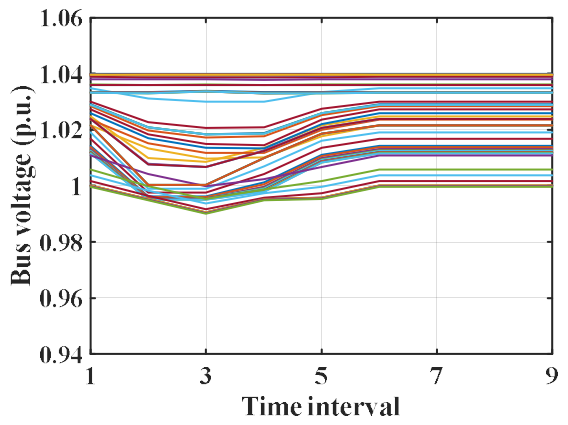

(a) Bus voltages

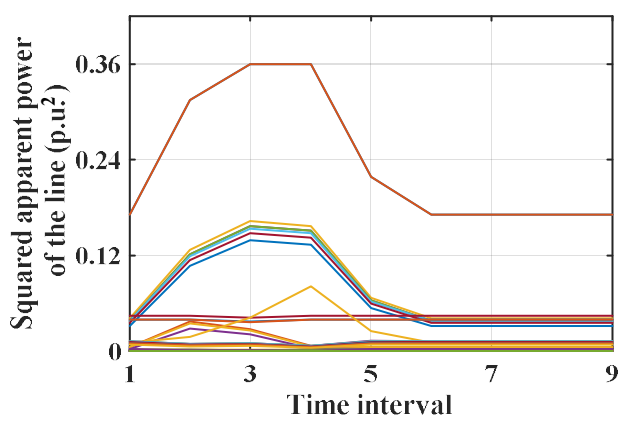

(b) Squared apparent power of the line

Fig. 13. DN operation state under the equilibrium state.

## VI. Conclusion

Considerable EV charging demands strengthen the interdependence between transportation and power networks. This paper constructs a coupled networks model to simulate the time-varying traffic flow and power network states. The multi-period AC OPF and the DTA with departure time/route choices and the dynamic queuing model are utilized to obtain more accurate operating states of coupled networks. The charging load fluctuation toll and bus voltage deviation penalty are respectively considered in TN and DN to improve the voltage quality in the equilibrium operation state.

Numerical experiments demonstrate that the proposed model captures the coupled networks equilibrium state efficiently. The electricity prices and travel costs are reduced by considering the dynamic travel and charging activities. Moreover, the maximum voltage deviation and the voltage variation among different time intervals are significantly decreased, and the voltage quality issues are tackled. The interaction analyses verify that bi-directional influences between TN and DN result in more uniformly distributed charging loads and increase the operational risks of both systems simultaneously. Moreover, tests on large networks verify the correctness and extensibility.

The proposed model is expected to realize the dynamic forecasting and estimation of coupled networks and provide the basis for operational control. Various uncertain factors of coupled networks will be considered for risk and resilience analyses of both systems.

## References

[1] Y. Wang, W. Y. Szeto, K. Han, and T. L. Friesz, “Dynamic traffic assignment: A review of the methodological advances for environmentally sustainable road transportation applications,” *Transp. Res. Part B: Methodol.*, vol. 111, pp. 370-394, 2018

[2] H. Das, M. Rahman, S. Li, and C. Tan, “Electric vehicles standards, charging infrastructure, and impact on grid integration: A technological review,” *Renew. Sustain. Energy Rev.*, vol. 120, p. 109618, 2020.

[3] Y. Sun, P. Zhao, L. Wang, and S. M. Malik, “Spatial and temporal modeling of coupled power and transportation systems: A comprehensive review,” *Energy Convers. Econ.*, 2021.

[4] E. A. Grunditz and T. Thiringer, “Performance analysis of current BEVs based on a comprehensive review of specifications,” *IEEE Trans. Transp. Electrif.*, vol. 2, no. 3, pp. 270-289, 2016.

[5] S. Sun, Q. Yang, and W. Yan, “Optimal temporal-spatial PEV charging scheduling in active power distribution networks,” *Protect. Control Mod. Power Syst.*, vol. 2, no. 1, pp. 1-10, 2017.

[6] R. Boudina, J. Wang, M. Benbouzid, F. Khoucha, and M. Boudour, “Impact evaluation of large scale integration of electric vehicles on power grid,” *Frontiers Energy*, vol. 14, no. 2, pp. 337-346, 2020.

[7] Z. Huang, B. Fang, and J. Deng, “Multi-objective optimization strategy for distribution network considering V2G-enabled electric vehicles in building integrated energy system,” *Protect. Control Mod. Power Syst.*, vol. 5, no. 1, pp. 1-8, 2020.

[8] Z. Ding, F. Teng, P. Sarikprueck, and Z. Hu, “Technical Review on Advanced Approaches for Electric Vehicle Charging Demand Management, Part II: Applications in Transportation System Coordination and Infrastructure Planning,” *IEEE Trans. Ind. Appl.*, vol. 56, no. 5, pp. 5695-5703, 2020.

[9] F. He, D. Wu, Y. Yin, and Y. Guan, “Optimal deployment of public charging stations for plug-in hybrid electric vehicles,” *Transp. Res. Part B: Methodol.*, vol. 47, pp. 87-101, 2013.

[10] W. Wei, L. Wu, J. Wang, and S. Mei, “Network equilibrium of coupled transportation and power distribution systems,” *IEEE Trans. Smart Grid*, vol. 9, no. 6, pp. 6764-6779, 2017

[11] W. Yao et al., “A multi-objective collaborative planning strategy for integrated power distribution and electric vehicle charging systems,” *IEEE Trans. Power Syst.*, vol. 29, no. 4, pp. 1811-1821, 2014.

[12] Y. Xiang, J. Liu, R. Li, F. Li, C. Gu, and S. Tang, “Economic planning of electric vehicle charging stations considering traffic constraints and load profile templates,” *Appl. Energy*, vol. 178, pp. 647-659, 2016.

[13] W. Wei, L. Wu, J. Wang, and S. Mei, “Expansion planning of urban electrified transportation networks: A mixed-integer convex programming approach,” *IEEE Trans. Transp. Electr.*, vol. 3, no. 1, pp. 210-224, 2017.

[14] W. Wei, S. Mei, L. Wu, M. Shahidehpour, and Y. Fang, “Optimal traffic-power flow in urban electrified transportation networks,” *IEEE Trans. Smart Grid*, vol. 8, no. 1, pp. 84-95, 2016.

[15] W. Wei, W. Danman, W. Qiuwei, M. Shafie-Khah, and J. P. Catalao, “Interdependence between transportation system and power distribution system: A comprehensive review on models and applications,” *J. Mod. Power Syst. Clean Energy*, vol. 7, no. 3, pp. 433-448, 2019.

[16] Y. Sun, Z. Chen, Z. Li, W. Tian, and M. Shahidehpour, “EV charging schedule in coupled constrained networks of transportation and power system,” *IEEE Trans. Smart Grid*, vol. 10, no. 5, pp. 4706-4716, 2018.

[17] S. Lv, Z. Wei, G. Sun, S. Chen, and H. Zang, “Optimal power and semi-dynamic traffic flow in urban electrified transportation networks,” *IEEE Trans. Smart Grid*, vol. 11, no. 3, pp. 1854-1865, 2019.

[18] Z. Zhou, X. Zhang, Q. Guo, and H. Sun, “Analyzing power and dynamic traffic flows in coupled power and transportation networks,” *Renew. Sustain. Energy Rev.*, vol. 135, pp. 110083, 2021.

[19] G. Sun, G. Li, S. Xia, M. Shahidehpour, X. Lu, and K. W. Chan, “ALADIN-Based Coordinated Operation of Power Distribution and Traffic Networks with Electric Vehicles,” *IEEE Trans. Ind Appl*, vol. 56, no. 5, pp. 5944-5954, 2020.

[20] S. W. Xie, Z. J. Hu and Y. Y. Wang, “Dynamic Flow Equilibrium of Urban Power and Transportation Networks Considering the Coupling in Time and Space,” *Proc Chin Soc Elect Eng*, 2021. [Online]. Available: https://kns.cnki.net/kcms/detail/11.2107.TM.20210226.1121.005.html

[21] S. Xie, Y. Xu and X. Zheng, “On Dynamic Network Equilibrium of a Coupled Power and Transportation Network,” *IEEE Trans. Smart Grid*, vol. 13, no. 2, pp. 1398-1411, 2021.

[22] R. Xie, W. Wei, M. E. Khodayar, J. Wang, and S. Mei, “Planning fully renewable powered charging stations on highways: A data-driven robust optimization approach,” *IEEE trans. Transp. Electr.*, vol. 4, no. 3, pp. 817-830, 2018.

[23] W. Wei, S. Mei, L. Wu, J. Wang, and Y. Fang, “Robust operation of distribution networks coupled with urban transportation infrastructures,” *IEEE Trans. Power Syst.*, vol. 32, no. 3, pp. 2118-2130, 2016.

[24] F. Xia, H. Chen, M. Shahidehpour, W. Gan, M. Yan, and L. Chen, “Distributed Expansion Planning of Electric Vehicle Dynamic Wireless Charging System in Coupled Power-Traffic Networks,” *IEEE Trans. on Smart Grid*, 2021.

[25] B. Ran and D. Boyce, Dynamic urban transportation network models: theory and implications for intelligent vehicle-highway systems. *Springer Science & Business Media*, 2012.

[26] W. Wei, L. Wu, J. Wang, and S. Mei, “Network equilibrium of coupled transportation and power distribution systems,” *IEEE Trans. Smart Grid*, vol. 9, no. 6, pp. 6764-6779, 2017.

[27] E. Beale and J. J. Forrest, “Global optimization using special ordered sets,” *Math. Program.*, vol. 10, no. 1, pp. 52-69, 1976.

[28] M. Farivar and S. H. Low, “Branch flow model: Relaxations and convexification-Part I,” *IEEE Trans. Power Syst.*, vol. 28, no. 3, pp. 2554-2564, 2013.

[29] L. J. LeBlanc, E. K. Morlok and W. P. Pierskalla, “An efficient approach to solving the road network equilibrium traffic assignment problem,” *Transp. Res.*, vol. 9, no. 5, pp. 309-318, 1975.

[30] M. E. Baran and F. F. Wu, “Optimal capacitor placement on radial distribution systems,” *IEEE Trans. Power Del.*, vol. 4, no. 1, pp. 725-734, 1989.